\documentclass[twocolumn,pre,showpacs,amsmath,longbibliography]{revtex4-1}
\usepackage{graphicx}
\usepackage{url}
\usepackage{cases}
\usepackage{bm}
\usepackage{color}
\usepackage[version=3]{mhchem}
\usepackage[normalem]{ulem}
\usepackage{hyperref}

\newcommand{\bra}{\left\langle}
\newcommand{\ket}{\right\rangle}

\newcommand{\pder}[2]{\frac{\partial #1}{\partial  #2}}
\newcommand{\pderf}[3]{\left(\frac{\partial #1}{\partial  #2}\right)_{#3}}

\newcommand{\sbkt}[1]{\langle#1\rangle}
\newcommand{\bbkt}[1]{\bigl\langle#1\bigr\rangle}
\newcommand{\Bbkt}[1]{\Bigl\langle#1\Bigr\rangle}

\newcommand{\kB}{k_{\rm B}}

\newcommand{\ep}{\epsilon}
\newcommand{\eq}{\mathrm{eq}}
\newcommand{\ex}{\mathrm{ex}}

\newcommand{\id}{\mathrm{ideal}}
\newcommand{\mix}{\mathrm{mix}}
\newcommand{\isotope}{\mathrm{}}

\newcommand{\subL}{\mathrm{L}}
\newcommand{\subR}{\mathrm{R}}

\newcommand{\subA}{\mathrm{A}}
\newcommand{\subB}{\mathrm{B}}
\newcommand{\subAB}{\mathrm{AB}}

\newcommand{\intt}{\int_{0}^{\tau} \!\! ds~}

\newcommand{\NCn}{\frac{N!}{n!(N-n)!}}

\begin{document}
\title{Work relation for determining the mixing free energy of small-scale mixtures}

\author{Akira Yoshida}
\email{a.yoshida.phys@gmail.com}
\affiliation {Department of Physics,  Ibaraki University, Mito 310-8512, Japan}
\author{Naoko Nakagawa}
\email{naoko.nakagawa.phys@vc.ibaraki.ac.jp}
\affiliation {Department of Physics,  Ibaraki University, Mito 310-8512, Japan}
\date{\today}


\begin{abstract}
In thermodynamically characterizing a mixture comprising a finite number of molecules, we consider two kinds of protocol for producing a mixture from a pure substance. The first is a single alchemical operation, whereas the second is a series of processes with feedback control in information thermodynamics and conventional mixing with semipermeable membranes. A comparison of the two numerically determined free-energy changes provides a combinatorial factor that indicates the indistinguishability of the molecules and an alternative Jarzynski equality. The comparison also uncovers a work relation for determining the mixing free energy without using semipermeable membranes. We demonstrate a numerical calculation of applying the work relation to a mixture of argon and krypton. The mixing free energy clearly shows the characteristics of liquid--vapor transition. 
\end{abstract}

\maketitle

\setcounter{tocdepth}{3}

\section{Introduction}

Solutions exhibit a variety of fascinating phenomena. Various combinations of solutes and solvents have been explored to create properties useful in scientific and industrial applications. Free energy and entropy are central quantities that characterize the properties of solutions \cite{barrow1979physical,landau1980statistical}.  
Thermodynamic measurements have been performed intensively to quantitatively determine these quantities, and the accumulated results have been integrated into a huge database \cite{NIST}.
In small-scale solutions, thermodynamic measurement is in its early stage of development \cite{yu2017review,harada2005equality, toyabe2011thermodynamic, ariga2018nonequilibrium} although
 recent micro-manipulation techniques have shed light on the non-triviality of small scales, 
including the anomalous diffusion of macromolecules and stabilization of protein folding by aggregation \cite{cheung2005molecular, squires2010fluid, wang2012brownian, chubynsky2014diffusing}. 
In cell sciences, liquid--liquid phase separation with the coexistence of  dilute and concentrated solutions has been intensively studied from the viewpoint of biological functions \cite{kroschwald2017gel, uversky2017intrinsically, dolgin2018lava, franzmann2018phase, bolognesi2019mutational, alberti2019liquid}.
Thermodynamic quantification is necessary to understand such interesting phenomena.
Numerical experiments may be powerful for the study of small systems.
This paper thus proposes an effective numerical method for the thermodynamic measurement of small-scale solutions.

Thermodynamics of small systems was proposed  in the $1960$'s on the basis of statistical mechanics 
\cite{hill1962thermodynamics}. 
In the past two decades, stochastic thermodynamics has been studied intensively, aiming for a physical understanding of molecular machines \cite{seifert2012stochastic,evans1993probability,jarzynski1997nonequilibrium, crooks2000path}. 
The change in free energy has been formulated as the Jarzynski or Crooks work relation
consistently with the second law of thermodynamics \cite{jarzynski1997nonequilibrium, crooks2000path}.
Using these relations, the change in free energy for the binding of biomolecules is determined by micro-manipulation \cite{smith1996overstretching,collin2005verification}.
Furthermore, information thermodynamics has been formulated by combining stochastic thermodynamics and information theory \cite{sagawa2010generalized,parrondo2015thermodynamics}, allowing us to approach biological phenomena from the perspective of information processing \cite{ito2015maxwell}.

For the numerical calculation of free energy, several methods have been and are being developed \cite{chipot2007free,cheng2018computing}.
Recently, alchemical free energy calculation is often used in the numerical study of biomolecules and drug discovery \cite{kollman1993free, kollman1996advances, simonson2002free, mobley2006use, mobley2012perspective, steinbrecher2017predicting, kuhn2020assessment, scheen2020hybrid},  which is an extension of the Kirkwood’s charging formula for determining chemical potential \cite{kirkwood1935}. 
The change in free energy is measured from the work required to substitute some parts of a large molecule alchemically; i.e., by changing microscopic parameters of the molecule.
Extending the idea, 
we may create a solution from a pure substance.
We then ask if the work relations are applicable to estimate the free energy of the solution.
We face two problems. 
The first problem is the indistinguishability of molecules. To create a solution alchemically, we need to choose some molecules to be manipulated from the indistinguishable molecules. Such a procedure is not involved in the usual alchemical method because it is designed for a single molecule. The problem may be related to the validity of the factorial in classical statistical mechanics adopted by Gibbs to recover extensivity \cite{gibbs1902, vankampen1984}. We here note that $\ln N!$ is asymptotically equal to $\ln N^N$ and therefore these two quantities are not distinguished in the thermodynamic limit. When dealing with small-scale solutions, their difference may appear.

The second problem is the quantity to be determined. The important quantity is the mixing free energy rather than the free energy for solutions. The mixing free energy, which involves excess chemical potentials or activity coefficients, corresponds to the work required for quasistatic mixing. It determines properties of a solution, such as equilibrium constants and solubility. However, theories for estimating the mixing free energy are limited to rather dilute solutions \cite{skyner2015review,kohns2016solvent,Debye_1923}.
A simpler numerical method applicable to the general concentration and valid regardless of the system size would be valuable. We thus propose a method for molecular dynamics simulations that estimates the mixing entropy of finite-size systems by combining the alchemical method with stochastic thermodynamics and information thermodynamics.

This paper is organized as follows. 
In Sec. \ref{s:setup}, we describe the setup of the system. 
In Sec. \ref{s:problem}, 
we address the problem of the conventional work relations when creating a solution from a pure substance
and propose \eqref{e:F_AB} as an alternative Jarzynski equality for determining the free energy of the solution.
Section \ref{s:protocol-alchemy} is devoted to showing \eqref{e:F_AB}. 
We obtain theoretically \eqref{e:Cthermo} and numerically Figs. \ref{fig:CvsCstat} from which we lead to combinatorial factor in \eqref{e:F_AB}.
We then proceed to the second part of the Paper.
In Sec. \ref{s:Gmix-iso}, we formulate a work relation for determining the mixing Gibbs free energy of isotope mixtures and  in Sec. \ref{s:Gmix-real}  extend the relation to general mixtures. The result is \eqref{e:Gmix}, which is estimated as \eqref{e:Gmix-N} or \eqref{e:Gmix-qs}.
Using \eqref{e:Gmix-qs} in a molecular dynamics simulation, we determine the mixing Gibbs free energy for a mixture of argon and krypton in Sec. \ref{s:Ar-Kr}. The result clearly shows the characteristics of the liquid--vapor transition.
Section \ref{s:concluding} is devoted to concluding remarks. 
All details of the model and the protocols for numerical examination are described in Appendices \ref{ap:model} and \ref{ap:protocol}.
The numerically determined free energies are examined carefully in Appendices \ref{ap:rho_v} and \ref{ap:Fii}. Appendix \ref{ap:statmech} compares our results with those of statistical mechanics.
Parameters for the numerical experiments in Sec. \ref{s:Ar-Kr} are specified in Appendix \ref{ap:Ar-Kr}.

\section{Setup} \label{s:setup}

We deal with classical systems of $N$ molecules packed in a rectangle container of volume $V$.
The container may be spatially partitioned by walls or semipermeable membranes.
The walls or membranes are rigid and transparent to heat and their positions do not fluctuate.
The surrounding environment is at a constant temperature $T$.
For simplicity, we limit the type of molecule to be monoatomic in this paper, but our proposed methods can be extended to more general molecules as discussed in Sec.~\ref{s:concluding}.
We write the Hamiltonian of the system as
\begin{align}
H(\Gamma;\bm{\alpha})=\sum_{i=1}^N \frac{\bm{p}_i^2}{2 m_i}+\Phi(\left\{\bm{r}_i\right\};\bm{\alpha}_{\Phi}),
\label{e:H-v}
\end{align}
where $\Gamma=(\left\{\bm{r}_i\right\}, \left\{ \bm{p}_i\right\})$ with the position $\bm{r}_i$ and momentum $\bm{p}_i$ for the $i$th molecule, and $(\left\{a_i\right\})$ is an abbreviation of $(a_1, a_2,\cdots, a_N)$. 
$m_i$ is the mass of the $i$th molecule and the potential $\Phi$ comprises the interaction among molecules and the interaction between molecules and walls or membranes of the container, which are parameterized by the set $\bm{\alpha}_{\Phi}$.
$\bm{\alpha}$ is the set of parameters in the Hamiltonian, $\bm{\alpha}=(\{m_i\}, \bm{\alpha}_{\Phi})$.
See Appendix \ref{ap:model} for an example of $\bm{\alpha}$ and $H(\Gamma;\bm{\alpha})$.

Suppose that an external operator changes the value of $\bm \alpha$   in the period $0\le t\le \tau$.
For a protocol $\hat{\bm{\alpha}}=(\bm{\alpha}(t))_{t\in[0,\tau]}$,
where ${\bm \alpha}_0={\bm \alpha}(0)$ and ${\bm \alpha}_1={\bm \alpha}(\tau)$,
the work done by the external operator is written as
\begin{align}
\hat W(\hat\Gamma)=\intt \frac{d\bm{\alpha}}{ds}\cdot
\left.\pder{H(\Gamma(s);\bm{\alpha})}{\bm{\alpha}}\right|_{\bm{\alpha}=\bm{\alpha}(s)}.
\label{e:work}
\end{align}
where $\hat \Gamma=(\Gamma(t))_{t\in [0,\tau]}$ is a trajectory in the phase space.
We assume below that the system is in equilibrium at ${\bm \alpha}_0$ for $t\le 0$.

\section{Problems of the work relation in microscopic operations} \label{s:problem}

We first consider macroscopic operations such as changing the volume of the container and the positions of  membranes.
Thermodynamic work corresponds to an ensemble average of the work over trajectories $\hat\Gamma$, which we write as $\langle\hat W\rangle$.
The difference in the Helmholtz free energy satisfies
\begin{align}
\Delta F\le \bbkt{\hat W},
\label{e:Helmholtz}
\end{align}
where $\Delta F=F(T,{\bm \alpha}_1,N)-F(T,{\bm \alpha}_0,N)$.
The equality holds in the quasistatic limit $\tau\rightarrow \infty$.
The work relation \eqref{e:Helmholtz} is reformulated as the Jarzynski equality \cite{jarzynski1997nonequilibrium}
\begin{align}
\Delta F=-\kB T\ln\bbkt{e^{-\beta \hat W}},
\label{e:Jarzynski}
\end{align}
where $\beta=(\kB T)^{-1}$ with the Boltzmann constant $\kB$.
With \eqref{e:Jarzynski}, the free energy becomes measurable  in mesoscopic systems of finite $N$ applicable to  single-molecule manipulations,
and moreover, the free-energy change can be identified from finite speed operations regardless of whether the system reaches equilibrium at $t=\tau$.

We next consider
microscopic operations called alchemical processes, which changes the attributes of molecules, such as the mass and size.
Alchemical methods are usually used to estimate the effect of substituting some groups into a large single molecule \cite{kollman1993free, kollman1996advances}.
We note that an alchemical method itself is not necessarily limited to single molecules but can be applied to multi-molecule systems so as to create a mixture from a pure substance.
Our first question is then whether the work relation \eqref{e:Jarzynski} can be used in determining 
the Helmholtz free energy of the mixture created from the pure substance.

The important quantity that determines the thermodynamic properties for the mixture is the mixing free energy $\Delta_\mix G$ rather than the free-energy difference between the mixture and the pure substance.
This corresponds to the work required for quasistatic mixing at constant temperature and constant pressure, which is the sum of the mixing entropy $\Delta_\mix S$ and the enthalpy change in mixing.
The mixing free energy gives the equilibrium constant and can be used as a variation function with which to identify the equilibrium state through its minimization.
We then ask the second question of whether there exists a work relation that can be used to determine $\Delta_\mix G$.

\section{Work relation for microscopic operations}\label{s:protocol-alchemy}

We consider a mixture of two species $\subA$ and $\subB$, whose numbers of molecules are $n$ and $N-n$, respectively.
We write the Helmholtz free energy for a pure substance of $\subA$ as $F_\subA(T,V,N)$ and that for the mixture as $F_\subAB(T,V,n,N-n)$.
To answer the first question, we focus on the free-energy difference between these two
\begin{align}
\Delta F\equiv F_\subAB(T,V,n,N-n)-F_\subA(T,V,N),
\label{e:F_AB-def}
\end{align}
and ask what is the work relation that can be used to obtain $\Delta F$ in the alchemical process.

\begin{figure}[bt]
\begin{center}
\includegraphics[scale=0.36]{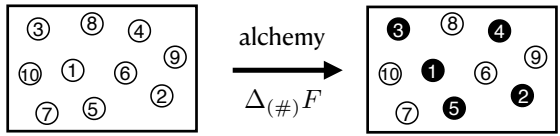}
\caption{Schematic figure of the alchemical process for creating the mixture of distinguished molecules to determine $\Delta_{(\#)}F$ in \eqref{e:Falc}. }
\label{fig:alchemy}
\end{center}
\end{figure}

We put $N$ molecules of species $\subA$ in a container and index all $N$ molecules in order; i.e.,  $i=1,2,\cdots, N$.
After relaxing the system to equilibrium, we change the attributes of the first $n$ molecules, $1\le i\le n$,  alchemically
as they become another species $\subB$ as depicted in Fig.~\ref{fig:alchemy}.
The resulting system is similar to a typical two-component mixture except that all molecules are indexed.
We call the molecules a distinguished mixture.
Although real molecules are never indexed,
the present procedure may be useful in numerical experiments.

We write the work for completing the alchemical process as $\hat W_{(\#)}(\hat\Gamma)$.
Applying \eqref{e:Jarzynski}, we can determine the free energy $F_{\subAB^\#}(T,V,n,N-n)$ for the distinguished mixture of $\subA$ and $\subB$ as 
\begin{align}
&\Delta_{(\#)} F \equiv F_{\subAB^\#}(T,V,n,N-n)-F_\subA(T,V,N), \\
&\Delta_{(\#)} F=-\kB T\ln \bbkt{e^{-\beta \hat W_{(\#)}}}.
\label{e:Falc}
\end{align}
In the subsection below, we show numerically that $\Delta F\neq \Delta_{(\#)} F$; i.e.,
\begin{align}
F_\subAB(T,V,n,N-n)\neq F_{\subAB^\#}(T,V,n,N-n),
\label{e:diffF}
\end{align}
which means that the alchemical process in Fig.~\ref{fig:alchemy} does not provide the free energy for the mixture.
From the thermodynamic argument in the following subsections with numerical examinations, we conclude that the formula that gives the true free-energy change \eqref{e:F_AB-def} is
\begin{align}
\Delta F=-\kB T\ln \left[\frac{N!}{n!(N-n)!}\bbkt{e^{-\beta \hat W_{(\#)}}}\right],
\label{e:F_AB}
\end{align}
which we propose as the Jarzynski work relation valid for general alchemical processes.
Applying the Jensen's inequality,  \eqref{e:F_AB} is written as
\begin{align}
\Delta F\le \bbkt{\hat W_{(\#)}}
-\kB T\ln \left[\frac{N!}{n!(N-n)!}\right].
\label{e:F_AB-qs}
\end{align}
The equality holds in the quasistatic limit.
The combinatorial factor in \eqref{e:F_AB} may seem natural considering the treatment of the indistinguishability of molecules in statistical mechanics. However, from a thermodynamic point of view, the combinatorial factor remains open. 
Below, we will numerically show the validity of \eqref{e:F_AB} using thermodynamic measurements.

\begin{figure}[tb]
\begin{center}
\includegraphics[scale=0.5]{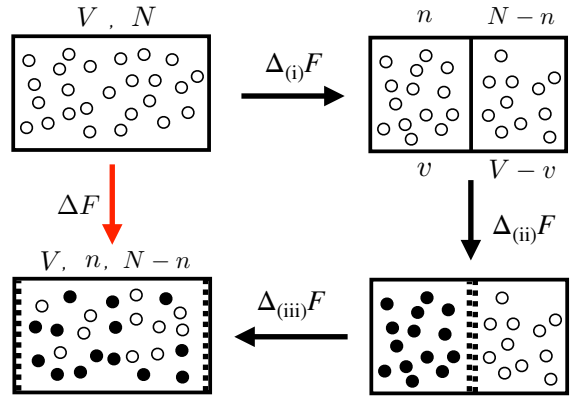}
\caption{Operation protocol for obtaining $F_{\subAB}$ by thermodynamic processes (i), (ii), and (iii). Refer to Table \ref{t:protocol} for a summary of respective processes. The red arrow indicates not an experimental protocol but a difference between a pure substance $\subA$ and a mixture of $\subA$ and $\subB$. Process (i) requires a feedback control to fix the number $n$ of molecules in the right chamber. }
\label{fig:protocol}
\end{center}
\end{figure}

\begin{table*}[tb]
\begin{center}
\begin{tabular}{|c|c|c|} \hline
  process  & operation  & $\Delta F$ \\ \hline \hline
      (i) & partition with feedback control & $F_\subA(T,v,n)+F_\subA(T,V-v,N-n)-F_\subA(T,V,N)$  \\ \hline
    (ii) & 
    alchemy the left side molecules
     & $F_\subB(T,v,n)-F_\subA(T,v,n)$ \\ \hline
    (iii) &mix by shifting the membranes   & $F_\subAB(T,V,n,N-n)-F_\subA(T,V-v,N-n)-F_\subB(T,v,n)$\\ \hline
    (\#) &    alchemy the molecules $1\le i\le n$
     & $F_{\subAB^\#}(T,V,n,N-n)-F_\subA(T,V,N)$ \\ \hline

  \end{tabular}
  \end{center}
  \caption{Summary of processes schematically illustrated in Figs.~\ref{fig:alchemy} and \ref{fig:protocol}. We use the subscripts $\subA$, $\subB$, and $\subAB$ to specify the quantities for the pure substance $\subA$, the pure substance $\subB$, and their mixture, respectively.}
\label{t:protocol}
\end{table*}

To show \eqref{e:F_AB},  we design a protocol comprising three processes
as schematically illustrated by the three black arrows in Fig.~\ref{fig:protocol} and summarized in Table \ref{t:protocol}.
Process (i) is the insertion of a wall partition that divides the container into two. This insertion is performed under feedback control as explained in Sec. \ref{s:process-i}, and the corresponding work relation is therefore given by information thermodynamics \cite{sagawa2010generalized, parrondo2015thermodynamics}.
In process (ii), we change the species of all molecules in the left chamber
and replace the wall partition with two ideal semi-permeable membranes.
We then shift each semi-permeable membrane to mix the two substances in process (iii).
We specify respective formulas of the work relation in Sec. \ref{s:process-ii} and Sec. \ref{s:process-iii}.

With these respective formulas, we numerically determine the respective free-energy changes $\Delta_{\rm (i)} F$, $\Delta_{\rm (ii)} F$ and  $\Delta_{\rm (iii)} F$.
The total change should be $\Delta F$ in \eqref{e:F_AB-def}:
\begin{align}
\Delta F=\Delta_{\rm (i)}F+\Delta_{\rm (ii)}F+\Delta_{\rm (iii)}F.
\label{e:F_AB-cycle}
\end{align}
The right-hand side of \eqref{e:F_AB-cycle} is thermodynamically definite
without the difficulty of the distinguishability of molecules.
In  Sec. \ref{s:verification}, we compare $\Delta F$ with the alchemical free-energy change $F_{\subAB^\#}-F_\subA$ determined according to \eqref{e:Falc}, which concludes the validity of \eqref{e:F_AB} within numerical error.

\subsection{Formula for $\Delta_{\rm (i)} F$}
\label{s:process-i}

In process (i), we spontaneously partition the container into two by inserting a rigid wall with a negligible thickness, where the volumes of the left and right chambers are $v$ and $V-v$.
Let $\Gamma$ be the microstate at the time of partitioning and 
 $n_{v}(\Gamma)$ be the number of molecules in the left chamber of volume $v$.
The probability distribution $\rho_{v}(n)$ for the number $n$ of molecules in the left chamber is written as
\begin{align}
\rho_{v}(n;T,V,N)= \int d\Gamma~ \delta_{n_{v}(\Gamma),n} \rho_\eq(\Gamma),
\label{e:rho_v}
\end{align}
where $\rho_\eq(\Gamma)$ is the canonical distribution before the partition and $\delta_{i,j}$ is the Kronecker delta. 

We perform feedback control to  insert a wall only when $n_{v}(\Gamma)=n$.
The work required for the spontaneous insertion depends on $\Gamma$, which we write as $\hat W_{\rm (i)}(\Gamma)$.
The change in the free energy satisfies
\begin{align}
e^{-\beta\Delta_{\rm (i)} F}=\int d\Gamma~ \delta_{n_{v}(\Gamma),n} e^{-\beta \hat W_{\rm (i)}(\Gamma)}
\rho_\eq(\Gamma).
\label{e:Jar-fb}
\end{align}
The relation \eqref{e:Jar-fb} is a version of the Jarzynski work relation putting $\delta_{n_{v}(\Gamma),n}$ in the integral because we perform the work only when $n=n_{v}(\Gamma)$.
\eqref{e:Jar-fb} belongs to the generalized Jarzynski relation derived in information thermodynamics, which is formulated for general feedback controls \cite{sagawa2010generalized}.

We design the interaction between the inserted wall and each molecule such that 
\begin{align}
\hat W_{\rm (i)}(\Gamma)=0.
\label{e:zeroW}
\end{align} 
This can be satisfied when each molecule acts as a point of mass with respect to the inserted wall. 
We choose this particular setting because $\Delta F$, which is a quantity to be determined, 
is independent of the properties of the inserted wall.

Substituting \eqref{e:zeroW} into the relation, we obtain
\begin{align}
\Delta_{\rm (i)} F=-\kB T\ln\rho_{v}(n;T,V,N).
\label{e:F_i}
\end{align}
Hereafter, we abbreviate $\rho_v(n;T,V,N)$ as $\rho_v(n)$.
The validity of \eqref{e:F_i} is examined from the perspective of statistical mechanics in Appendix \ref{ap:statmech}.
We are now able to determine $\Delta_{\rm (i)}F$ without performing the insertion. 
One should simply count the number of molecules in the region corresponding to the left chamber from time to time and determine $\rho_{v}(n)$. 

Note that the protocol (i) is common regardless of the species of the initial pure substance or the composition of the final mixture.
$\Delta_{\rm (i)}F$ is expressed by a general form with an error of $o(\ln N)$. See \eqref{e:rho_v-apprx} and Appendix \ref{ap:rho_v}.

\subsection{Work relation for $\Delta_{\rm (ii)}F$}
\label{s:process-ii}

We next consider $\Delta_{\rm (ii)} F$, 
which corresponds to the difference between two pure substances as
\begin{align}
\Delta_{\rm (ii)}F=F_\subB(T,v,n)-F_\subA(T,v,n),
\end{align}
where $F_\subB$ is the Helmholtz free energy  for the substance $\subB$.
We apply an alchemical process to determine $\Delta_{\rm (ii)}F$.
We change all molecules in the left chamber to other species according to the same protocol.
We emphasize that we do not need to consider whether process (ii) is a microscopic or macroscopic operation
because all molecules in the left chamber are changed in the same manner.
We add that there are several numerical ways to determine $\Delta_{\rm (ii)}F$, and that the values of $F_\subA$ and $F_\subB$ may be referenced from a database \cite{NIST}.

The process (ii) is expressed by the change in the Hamiltonian, 
and we thus define the required work $\hat W_{\rm (ii)}$ for each trajectory according to \eqref{e:work}.
The free-energy change in the process (ii) is calculated using the usual Jarzynski equality \eqref{e:Jarzynski} as 
\begin{align}
\Delta_{\rm (ii)}F=-\kB T\ln\bbkt{e^{-\beta \hat W_{\rm (ii)}}}.
\label{e:F_ii}
\end{align}
$\Delta_{\rm (ii)} F$ corresponds to the free-energy change for the total system
because this process does not affect the free energy of the right chamber.
At constant volume, the pressure in the left chamber may be changed by process (ii)  and different from that of the right chamber in general.

Before proceeding to the next process, we replace the wall partition inserted in (i) with two semi-permeable membranes.
The thickness of the two membranes is the same as that of the wall partition. 
Assuming that the two membranes do not interact with each other,
the replacement does not require thermodynamic work or affect the free energy of the system.
We assume that each membrane is ideal as it does not interact with the other 
and allows one molecule species to pass without interaction but completely blocks the other molecule species from passing via a repulsive force.

\begin{figure}[tb]
\begin{center}
\includegraphics[scale=0.45]{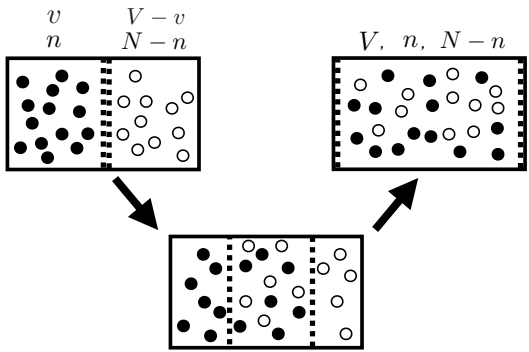}
\caption{Mixing process using two semipermeable membranes. $\Delta_{\rm (iii)} F$ is calculated from the work required to shift the two semipermeable membranes. }
\label{fig:mixing}
\end{center}
\end{figure}

\subsection{Work relation for $\Delta_{\rm (iii)}F$}
\label{s:process-iii}

Process (iii) corresponds to a standard mixing process of two pure substances, which appears in textbooks on thermodynamics \cite{gibbs1875on, fermi1956thermodynamics} for the demonstration of mixing entropy $\Delta_\mix S$ and mixing free energy $\Delta_\mix G$.
Upon shifting each semi-permeable membrane slowly as shown in Fig.~\ref{fig:mixing},  the two substances mix together between the two membranes. When each membrane reaches the left or right boundary wall,  the container is filled with the mixture of the two substances.

Before we go further, we emphasize that this mixing process makes for a difficult computation involving huge computational resources.
This is because the speed of the shift of the membranes should be much lower than the velocity of molecules,
whereas the distance between each membrane and each boundary wall is macroscopic.
When we perform alchemical protocols such as the process in Fig.~\ref{fig:alchemy} and process (ii) in Fig.~\ref{fig:protocol}, the required time step is approximately $O(N^0)$ in calculating the work $\hat W$ per trajectory.
Meanwhile, process (iii) requires, at least, $O(N)$ time steps  per trajectory.
This number could be more, such as $O(N^2)$, because we need to relax the system close to equilibrium after each slight shift of the membranes even though we use the Jarzynski equality.
Such a high-cost calculation is hard to complete with a large enough system, and  this is likely the reason that process (iii) is not usually used in the numerical investigation of the mixing entropy.
Only when the system is as small as $N\le 100$ can we perform process (iii) in determining $\Delta_{\rm (iii)}F$ with good accuracy as described below.

Suppose the work required to shift a membrane to the left boundary is $\hat W_{\rm (iii)}^\subL(\hat\Gamma)$
and that required to shift another membrane to the right is $\hat W_{\rm (iii)}^\subR(\hat\Gamma)$.
The free-energy change in process (iii) is estimated to be
\begin{align}
\Delta_{\rm (iii)}F=-\kB T\ln\Bbkt{e^{-\beta \left(\hat W_{\rm (iii)}^\subL+\hat W_{\rm (iii)}^\subR\right)}}.
\label{e:F_iii}
\end{align}
As noticed for process (ii), 
the initial pressures may not be balanced between the left and right chambers in general,
and therefore,  process (iii) may not be quasistatic even in the thermodynamic limit.
Because we are adopting the Jarzynski work relation, such problems do not affect the estimate of the free-energy change.

\subsection{Difference between $F_{\subAB^\#}$ and $F_\subAB$} 

From \eqref{e:F_i}, \eqref{e:F_ii}  and \eqref{e:F_iii}, we obtain
\begin{align}
&\Delta F=
-\kB T\ln
\left[
\rho_{v}(n)
\bbkt{e^{-\beta \hat W_{\rm (ii)}}}
\bbkt{e^{-\beta (\hat W_{\rm (iii)}^\subL+\hat W_{\rm (iii)}^\subR)}}
\right],
\label{e:Ftot}
\end{align}
which is a thermodynamically valid formula for the free-energy difference \eqref{e:F_AB-def} between the pure substance $\subA$ to the mixture of $\subA$ and $\subB$.
We note that the left-hand side of \eqref{e:Ftot} does not depend on $v$, whereas the respective quantities on the right-hand side are determined for a given $v$.
Thus, the dependence on $v$ should be canceled out by multiplying  the three quantities.

Combining \eqref{e:Ftot} with \eqref{e:Falc}, we have 
\begin{align}
\beta(F_\subAB - F_{\subAB^\#})=-\ln
\frac{
\rho_{v}(n)
\bbkt{e^{-\beta \hat W_{\rm (ii)}}}
\bbkt{e^{-\beta (\hat W_{\rm (iii)}^\subL+\hat W_{\rm (iii)}^\subR)}}
}{\bbkt{e^{-\beta \hat W_{(\#)}}}}.
\label{e:Cthermo}
\end{align}
We emphasize that  the right-hand side of \eqref{e:Cthermo} comprises quantities measurable in numerical thermodynamic experiments.
By measuring these quantities numerically, we obtain an answer to the first question raised in Sec.  \ref{s:problem},
whether $F_\subAB=F_{\subAB^\#}$ or $F_\subAB\neq F_{\subAB^\#}$.

\begin{figure}
\begin{center}
\includegraphics[scale=0.58]{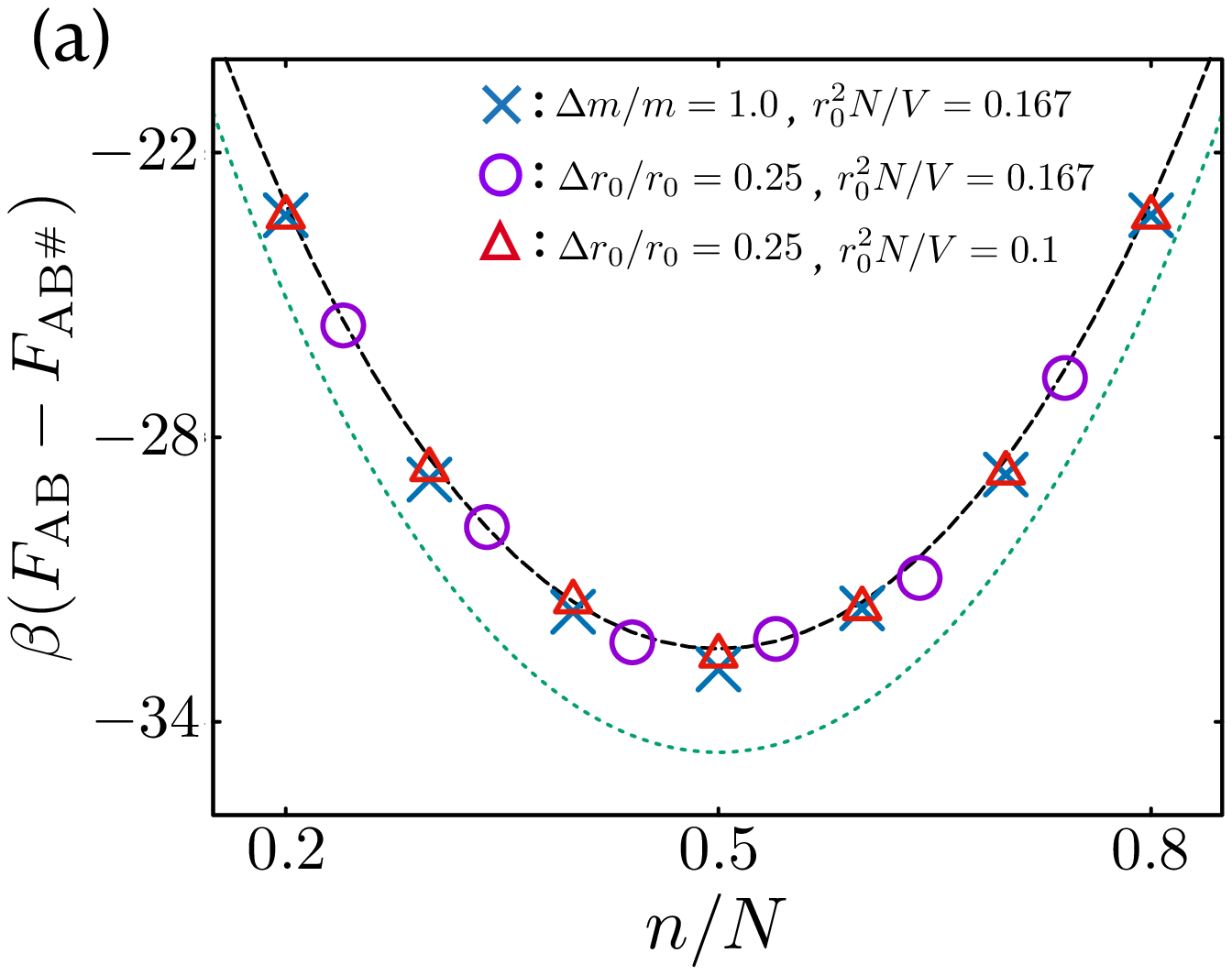}
\includegraphics[scale=0.6]{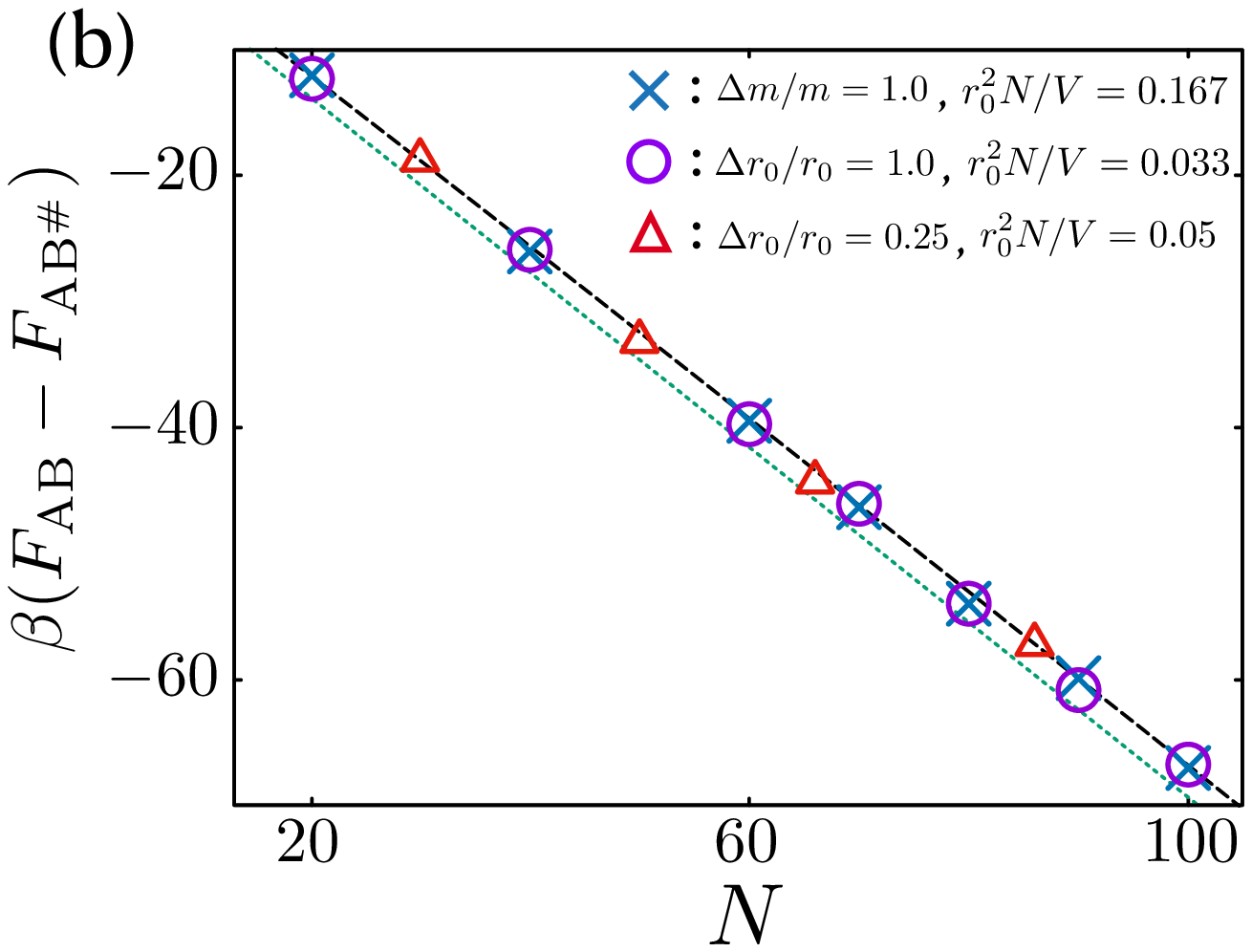}
\caption{
$\beta(F_\subAB - F_{\subAB^\#})$ as the right-hand side of \eqref{e:Cthermo} for three types of mixture: a mixture of isotopes with different masses ($\times$,  blue) and two mixtures of two types of molecule of different size ($\circ$,  purple and $\triangle$, red).  
The dashed black line shows  $-\ln[N!/n!(N-n)!]$ and 
the dotted green line shows $n\ln n/N+(N-n)\ln (N-n)/N$.
(a) $\beta(F_\subAB - F_{\subAB^\#})$ as a function of $n/N$ for $N=50$ with changing $n$
and (b) $\beta(F_\subAB - F_{\subAB^\#})$ as a function of $N$ when $n/N=0.5$.
Error bars are not plotted as they would be smaller than the data point. 
}
\label{fig:CvsCstat}
\end{center}
\end{figure}

\subsection{Numerical results for the right-hand side of \protect{\eqref{e:Cthermo}}} 
\label{s:verification}

To numerically estimate the right-hand side of \eqref{e:Cthermo}, we perform molecular dynamics simulations for two types of model mixtures.
One is a mixture comprising two isotopes of a monoatomic molecule, where the masses of the species $\subA$ and $\subB$ are set as $m$ and $m+\Delta m$, respectively.
The other is a mixture comprising monoatomic molecules that are different in size.
The radius of the species $\subA$ is $r_0$, whereas that of $\subB$ is $r_0+\Delta r_0$.
The details of the models are described in Appendix \ref{ap:model}, and the explicit protocols are specified in Appendix \ref{ap:protocol}.
Numerical results on $\Delta_{\rm (i)}F$, $\Delta_{\rm (ii)}F$, and $\Delta_{\rm (iii)}F$ are presented in Appendices \ref{ap:rho_v} and \ref{ap:Fii}.

Figures \ref{fig:CvsCstat} are simultaneous plots of the numerical results for the right-hand side of \eqref{e:Cthermo} for the two mixtures. 
We find that all data converge to a black dashed line; i.e.,
\begin{align}
\ln
\frac{
\rho_{v}(n)
\bbkt{e^{-\beta \hat W_{\rm (ii)}}}
\bbkt{e^{-\beta (\hat W_{\rm (iii)}^\subL+\hat W_{\rm (iii)}^\subR)}}
}{\bbkt{e^{-\beta \hat W_{(\#)}}}}
=\ln\frac{N!}{n!(N-n)!}
\label{e:distinguish}
\end{align}
over a wide range of $n/N$ and $N$ for fixed $V$ and $\beta$.
The volume $v$ for the left chamber is chosen as $v=Vn/N$ as it gives the most probable value for $n$. 
We emphasize that the plots in Figs.~\ref{fig:CvsCstat} contain the data for completely different mixtures and various values of
$\Delta m/m$ and $\Delta r_0/r_0$. 
Therefore, the convergence strongly suggests the universality of the functional form \eqref{e:distinguish}.
Combining \eqref{e:Cthermo} with \eqref{e:distinguish},  we conclude \eqref{e:diffF}; i.e., $F_\subAB \neq F_{\subAB^\#}$,
and more preciesely,
\begin{align}
\beta(F_\subAB-F_{\subAB^\#})=-\ln\frac{N!}{n!(N-n)!}.
\label{e:combination}
\end{align}
Combining \eqref{e:Falc} and \eqref{e:combination}, we obtain the Jarzynski equality \eqref{e:F_AB}.

The system sizes, $20\le N \le 100$, in the numerical experiments are small enough to distinguish $\ln N!$ from its asymptotic form $N\ln N$.
In Figs.~\ref{fig:CvsCstat},  we show a dotted line 
corresponding to the asymptotic of \eqref{e:combination},  $n\ln\frac{n}{N}+(N-n)\ln\frac{N-n}{N}$, 
estimated using Stirling's formula.
The dotted line does not coincide with the numerical results; i.e.,
\begin{align}
\beta(F_\subAB-F_{\subAB^\#})\neq-\ln\frac{N^{N}}{n^{n}(N-n)^{N-n}}.
\label{e:combination_2}
\end{align}

Once we obtain \eqref{e:F_AB}  and \eqref{e:combination}, we may recognize the combinatorial factor as the manifestation of the indistinguishability of molecules in statistical mechanics. Historically, the factorial $N!$ was introduced to statistical mechanics by Gibbs to recover extensivity \cite{gibbs1902, vankampen1984} and then became convincing owing to the consistency with quantum mechanics. However, the factorial $N!$ remains experimentally unverified in classical systems because $\ln N!$ is hardly distinguishable from $N \ln N$ in the macroscopic limit. The non-equality \eqref{e:combination_2} denies the possibility that $N \ln N$ is the factor to recover the extensivity in classical systems. Moreover, our numerical result suggests that one may derive the indistinguishability of molecules for classical systems by deriving the relation \eqref{e:distinguish} theoretically.

\section{Numerical method of calculating $\Delta_\mix G$ for an isotope mixture}\label{s:Gmix-iso}

When the pressure and volume are kept constant in process (iii), we have $\Delta_{\rm (iii)}F=\Delta_\mix G$.
Such a situation occurs for isotope mixtures as explained below, and we can calculate $\Delta_\mix G$
from the numerical scheme to use the relation \eqref{e:F_iii}.
This may be part of the  answer to the second problem raised in Sec. \ref{s:problem}.
However, we note that the calculation is rather impractical as discussed in the previous sections.
We thus propose another scheme to calculate $\Delta_\mix G$ without performing the macroscopic operations as process (iii).
In this section, we concentrate on a mixture of isotopes, which is a simpler example for finding a formula for $\Delta_\mix G$, and we then extend the method to other mixtures in the next section.

Below,
we limit $v$ as
\begin{align}
v=\frac{n}{N}V,
\end{align}
which gives  a natural choice of $n$ corresponding to the most probable value.

A mixture of isotopes comprises two substances different only in their mass. 
The interaction potential $\Phi(\left\{{\bm r}_i\right\})$ is common between the pure substances and the resulting mixture.
The pressure $p$ is kept constant over all processes shown in Table \ref{t:protocol} 
while the system is at constant volume.
It is thus possible to regard that all processes at constant volume are performed at constant pressure,
which results in $\Delta G=\Delta F$ for all processes.
Moreover,  because the internal energy of the system $U=\sbkt{H}$ never changes,
we have $\Delta S=-\Delta F/T$.
Process (iii) then corresponds to a usual mixing process with 
\begin{align}
&\Delta_\mix G^\isotope=\Delta_{\rm (iii)}F,\\
&\Delta_\mix S^\isotope=-\Delta_{\rm (iii)} F/T,
\end{align} 

We note that the processes in Fig.~\ref{fig:protocol} form a cycle 
once we identify the operation depicted by the red arrow, whose free-energy difference is given by \eqref{e:F_AB} with the alchemical operation in Fig.~\ref{fig:alchemy}. 
Thus, substituting \eqref{e:F_AB}, \eqref{e:F_i}, and \eqref{e:F_ii} into $\Delta_{\rm (iii)}F=\Delta F-\Delta_{\rm (i)}F-\Delta_{\rm (ii)}F$, we obtain the mixing Gibbs free energy for two isotopes as
\begin{align}
\Delta_\mix G^\isotope=-\kB T\ln \left[\frac{N!}{n!(N-n)!\rho_{v}(n)}
\frac{\bbkt{e^{-\beta \hat W_{(\#)}}}}{\bbkt{e^{-\beta \hat W_{\rm (ii)}}}}\right].
\label{e:Gmix-iso}
\end{align}
The right-hand side of \eqref{e:Gmix-iso} comprises numerically accessible quantities whose computational cost is much lower than the cost for performing \eqref{e:F_iii}.

The relation \eqref{e:Gmix-iso} is further simplified using
\begin{align}
\ln \rho_v(n=Nv/V)=-\frac{1}{2}\ln N+o(\ln N),
\label{e:rho_v-apprx}
\end{align}
which is derived in  Appendix  \ref{ap:rho_v}.
Combining the estimate \eqref{e:rho_v-apprx} with Stirling's formula,
$\ln N!=N\ln N-N+\frac{1}{2}\ln N +o(\ln N)$,
we have
\begin{align}
\ln \frac{N!}{n!(N-n)!\rho_{v}(n)}=\frac{\Delta_\mix S^{\id}}{\kB}+o(\ln N),
\label{e:S^id-apprx}
\end{align}
where $\Delta_\mix S^{\id} = -\kB [n\ln n/N+(N-n)\ln (N-n)/N ]$ is the mixing entropy for ideal solutions.
Substituting \eqref{e:S^id-apprx} into \eqref{e:Gmix-iso}, we arrive at
\begin{align}
\Delta_\mix G^\isotope=-T\Delta_{\rm mix} &S^{\id}-\kB T\ln 
\frac{\bbkt{e^{-\beta \hat W_{(\#)}}}}{\bbkt{e^{-\beta \hat W_{\rm (ii)}}}} +o(\ln N).
\label{e:Gmix-iso-fi}
\end{align}
Formula \eqref{e:Gmix-iso-fi} indicates that $\Delta_\mix G^\isotope$ is accessible only by the two alchemical processes.
Here, we comment that, in the case of isotopes,
the two ensemble averages in the second term of the right-hand side are always equal and $\Delta_\mix G^\isotope=-T\Delta_\mix S^{\id} + o(\ln N)$.
Furthermore, because the mixing enthalpy $\Delta_\mix H^\isotope=0$, we have
 \begin{align}
 \Delta_\mix S^\isotope= \kB \ln \frac{N^{N}}{n^n(N-n)^{N-n}} + o(\ln N).
 \label{e:Smix-iso}
 \end{align}
It is straightforward that
 \begin{align}
\Delta_\mix S^\isotope \neq \kB \ln \NCn.
\label{e:Smix-iso-notcombi}
\end{align}
Thus, the mixing entropy of isotopes does not correspond to the combinatorial entropy but rather behaves as the ideal mixing entropy $\Delta_\mix S^\id$ even at $N=20$ far from the thermodynamic limit.
For general mixtures, the functional form of the mixing entropy is not necessarily to be \eqref{e:Smix-iso}, whereas the formula \eqref{e:Gmix-iso-fi} remains valid. It provides a new method of obtaining the mixing free energy $\Delta_\mix G$ as explained in the next section.

\section{Generalization to real solutions}\label{s:Gmix-real}

We now extend the formulas \eqref{e:Gmix-iso} and \eqref{e:Gmix-iso-fi} from the mixture of isotopes to general real solutions.
In Sec. \ref{s:setup-p}, we set up the system and its Hamiltonian at constant pressure and explain the version of the Jarzynski work relation for constant pressure.
In Sec. \ref{s:Gmix-p}, we propose the formulas \eqref{e:Gmix}, \eqref{e:Gmix-N} and \eqref{e:Gmix-qs} for $\Delta_\mix G$  with two types of alchemical work $\hat W_{\rm (ii)}$ and $\hat W_{(\#)}$ and restate them as the relations for the activity coefficients in \eqref{e:act}.

\subsection{Setup at constant pressure}\label{s:setup-p}

When a certain wall of the container is replaced with a movable wall at constant pressure of $p$,
the system's Hamiltonian changes to
\begin{align}
H_p(\Gamma,V;\bm{\alpha})=H(\Gamma;\bm{\alpha})+pV.
\label{e:H-p}
\end{align}
Note that the pressure $p$ is a fixed constant,  whereas the volume $V$ becomes a microscopic variable for the Hamiltonian.
A trajectory in phase space is given by $(\hat\Gamma,\hat V)=(\Gamma(t),V(t))_{t\in[0,\tau]}$.
When a set of parameters $\bm \alpha$ is used, the required work is written as
\begin{align}
\hat W(\hat\Gamma,\hat V) 
&=\intt \frac{d\bm{\alpha}}{ds}\cdot\left.\pder{H(\Gamma(s);\bm{\alpha})}{\bm{\alpha}}\right|_{\bm{\alpha}=\bm{\alpha}(s)},
\end{align}
where $\hat W$ is determined by the Hamiltonian $H(\Gamma)$ and not by $H_p(\Gamma,V)$  because the second term $pV$ in \eqref{e:H-p} does not depend on $\bm\alpha$ at fixed $p$.
Similarly to the system at constant volume, 
the work in macroscopic operations or single-molecule manipulations leads to the Jarzynski work relation
\begin{align}
\Delta G=-\kB T\ln\bbkt{e^{-\beta \hat W}},
\label{e:Jarzynski-p}
\end{align}
where $\Delta G$ is the change in the Gibbs free energy $G(T,p, \bm{\alpha},N)$.
$\sbkt{\cdot}$ is the average over trajectories $(\hat\Gamma,\hat V)$ starting from equilibrium states at constant pressure,
which corresponds to the usual ensemble average in numerical experiments starting after a sufficient relaxation.

\subsection{Formulas for mixing free energy and activity coefficients}\label{s:Gmix-p}

\begin{figure}[bt]
\begin{center}
\includegraphics[scale=0.49]{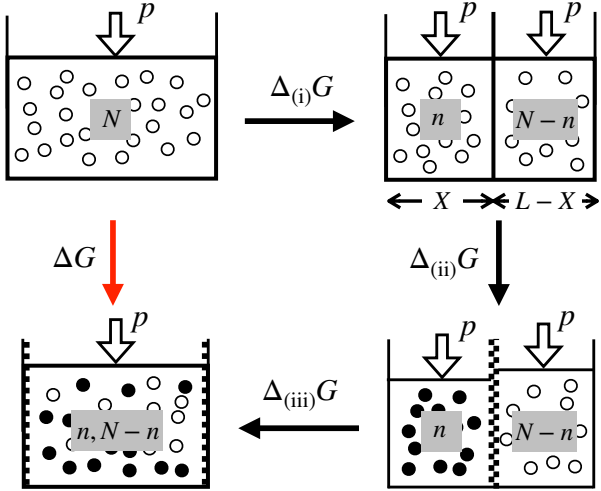}
\caption{Cycle at constant pressure designed in parallel to the cycle at constant volume in Fig.~\ref{fig:protocol}.
}
\label{fig:protocol-p}
\end{center}
\end{figure}

\begin{figure}[bt]
\begin{center}
\includegraphics[scale=0.38]{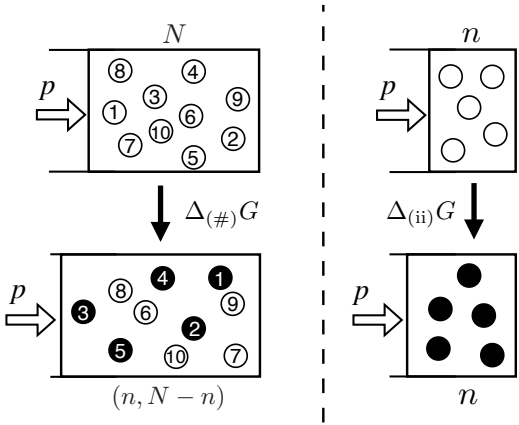}
\caption{Two alchemical processes to be calculated in the determination of $\Delta_\mix G$ using formula \eqref{e:Gmix}, \eqref{e:Gmix-N}, or \eqref{e:Gmix-qs}.
}
\label{fig:G-op}
\end{center}
\end{figure}

Following the previous argument, we consider a cycle at constant pressure shown in Fig.~\ref{fig:protocol-p}, which is similar to the cycle in Fig.~\ref{fig:protocol} at constant volume.

Performing all processes at constant pressure of $p$,
the cycle in Fig.~\ref{fig:protocol-p} leads to
\begin{align}
\Delta G=\Delta_{\rm (i)}G +\Delta_{\rm (ii)}G+\Delta_{\rm (iii)}G,
\label{e:G-cycle}
\end{align}
where
\begin{align}
\Delta G\equiv G_\subAB(T,p,n,N-n)-G_\subA(T,p,N).
\end{align}
Operationally, $\Delta G$ is the free-energy difference due to 
the alchemical process 
that changes the pure substance of $\subA$ into the mixture of $\subA$ and $\subB$ as depicted by the red arrow in Fig.~\ref{fig:protocol-p}.
Referring to formula \eqref{e:F_AB}, we expect that
\begin{align}
\Delta G=-\kB T\ln \left[\frac{N!}{n!(N-n)!}\bbkt{e^{-\beta \hat W_{(\#)}}}\right],
\label{e:G}
\end{align}
where $\hat W_{(\#)}$ is defined on the system of the distinguishable molecules according to the alchemical process illustrated in the left  figure of  Fig.~\ref{fig:G-op}.

To perform process (i) in Fig.~\ref{fig:protocol-p}, let $L$ be the length of the container and choose $X$ as the position to insert a wall partition. We then observe the number $n$ in the region of $x<X$ and define $\rho_X(n)$ in parallel to \eqref{e:rho_v} with the canonical distribution $\rho_\eq(\Gamma)$ at constant pressure. The change in Gibbs free energy for process (i) is formulated as
\begin{align}
\Delta_{\rm (i)}G=-\kB T\ln \rho_X(n)
\label{e:G_i}
\end{align}
similarly to \eqref{e:F_i}.
The estimate \eqref{e:rho_v-apprx} is also valid for the most probable value of $n$; i.e., $n=NX/L$.
Alchemical process (ii) at constant pressure is shown
 in the right figure of Fig.~\ref{fig:G-op}, where the right chamber is omitted.
 \eqref{e:Jarzynski-p} leads to
\begin{align}
\Delta_{\rm (ii)}G=-\kB T\ln\bbkt{e^{-\beta \hat W_{\rm (ii)}}}.
\label{e:G_ii}
\end{align}
Protocol (iii) at constant pressure is exactly the mixing process for the two pure substances $\subA$ and $\subB$; i.e.,
\begin{align}
\Delta_{\rm (iii)}G=\Delta_\mix G.
\label{e:G_iii}
\end{align}

Substituting \eqref{e:G}, \eqref{e:G_i}, \eqref{e:G_ii}, and \eqref{e:G_iii} into \eqref{e:G-cycle},
we have the formula for the mixing free energy as
\begin{align}
\Delta_{\rm mix}G
=
-\kB T\ln \left[\frac{N!}{n!(N-n)!\rho_{X}(n)}
\frac{\bbkt{e^{-\beta \hat W_{(\#)}}}}{\bbkt{e^{-\beta \hat W_{\rm (ii)}}}}\right].
\label{e:Gmix}
\end{align}
Obviously, formula \eqref{e:Gmix} for general mixtures is consistent with \eqref{e:Gmix-iso} for isotope mixtures, and it is therefore considered to be a general work relation giving the mixing free energy $\Delta_\mix G$.
Once we obtain \eqref{e:Gmix}, a similar transformation from \eqref{e:Gmix-iso} to \eqref{e:Gmix-iso-fi} is possible, which leads to
\begin{align}
\Delta_{\rm mix}G
=
-T\Delta_\mix S^{\id}-\kB T\ln\frac{\bbkt{e^{-\beta \hat W_{(\#)}}}}{\bbkt{e^{-\beta \hat W_{\rm (ii)}}}}+o(\ln N).
\label{e:Gmix-N}
\end{align}
Recalling \eqref{e:Falc},  the Gibbs free-energy change in the alchemical process for the distinguished molecules in the right figure of Fig.~\ref{fig:G-op} is written as
\begin{align}
\Delta_{(\#)}G=G_{\subAB^{\#}}-G_\subA=-\kB T\ln\sbkt{e^{-\beta \hat W_{(\#)}}}.
\label{e:Galc}
\end{align}
Substituting \eqref{e:G_ii} and \eqref{e:Galc}  into \eqref{e:Gmix-N}, we obtain 
\begin{align}
\Delta_\mix G=-T\Delta_\mix S^{\id}-\Delta_{\rm (ii)} G+\Delta_{(\#)} G+o(\ln N).
\label{e:Gmix-qs}
\end{align}
We emphasize that $\Delta_\mix G$ is determined just from two alchemical processes in Fig.~\ref{fig:G-op}.
Compared with the calculation of the mixing free energy along process (iii), the numerical cost to calculate \eqref{e:Gmix-N} or \eqref{e:Gmix-qs} is low.

Mixing changes the thermodynamic properties of each substance. 
This change is represented by excess chemical potential; i.e., the deviation of chemical potential from that of each pure substance.
Letting the concentration of the mixture be $c\equiv n/N$, the excess chemical potential is written as 
\begin{align}
&\beta\mu^\ex_\subA(T,p,c)=\ln c +\ln\gamma_\subA,\label{e:gammaA}\\
&\beta\mu^\ex_\subB(T,p,c)=\ln (1-c) +\ln\gamma_\subB,\label{e:gammaB}
\end{align}
with the activity coefficients $\gamma_\subA(T,p,c)$ and $\gamma_\subB(T,p,c)$.
When $\gamma_{\subA}=\gamma_\subB=1$, the mixture is ideal; i.e, a molecule of substance $\subA$ does not interact with a molecule of $\subB$. Therefore, the values of $\ln\gamma_\subA$ and $\ln\gamma_\subB$ represent the intrinsic properties of the mixture that result from the interaction of the two pure substances.
For the total mixture, the effect of mixing is summarized by the mixing Gibbs free energy $\Delta_\mix G$,
\begin{align}
\Delta_\mix G=n\mu_\subA^\ex+(N-n)\mu_\subB^\ex.
\label{e:Gmix-muex}
\end{align}
Summarizing \eqref{e:gammaA},  \eqref{e:gammaB}, and  \eqref{e:Gmix-muex}, we have
\begin{align}
\Delta_\mix G+T\Delta_\mix S^{\id}=\kB T\left[n\ln\gamma_\subA+(N-n)\ln\gamma_\subB \right].
\end{align}
Thus,  \eqref{e:Gmix-N} and \eqref{e:Gmix-qs} lead to a relation for the activity coefficients as
\begin{align}
&n\ln \gamma_\subA+(N-n)\ln\gamma_\subB
=-\ln \frac{\bbkt{e^{-\beta \hat W_{(\#)}}}}{\bbkt{e^{-\beta \hat W_{\rm (ii)}}}}+o(\ln N)\nonumber\\
&\qquad\qquad =\beta(\Delta_{(\#)}G-\Delta_{\rm (ii)}G)+o(\ln N).
\label{e:act}
\end{align}
The estimates of activity coefficients  are a major issue
in the research of mixtures, especially from the point of chemical engineering.
The relation \eqref{e:act} may offer a new method of estimating the activity coefficients for various mixtures and solutions, which involves only a molecular dynamics simulation with two types of alchemical process.

\section{Numerical demonstration of $\Delta_\mix G$ for a mixture of argon and krypton} \label{s:Ar-Kr}

\begin{figure}
\begin{center}
\includegraphics[scale=0.5]{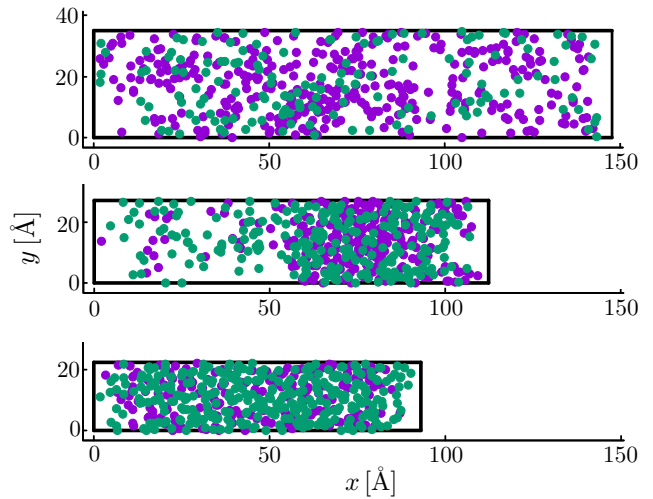}
\caption{Snapshots of the particle distribution for the mixture of argon (purple) and krypton (green). The three-dimensional space inside the container is projected onto the $xy$ plane.
Upper, middle, and bottom figures are for $c_{\rm Kr}=0.3$, $0.5$, and $0.65$, respectively.
The middle figure clearly shows the separation of liquid from vapor.}
\label{fig:Snap-RareGas}
\end{center}
\end{figure}

\begin{figure}
\begin{center}
\includegraphics[scale=0.6]{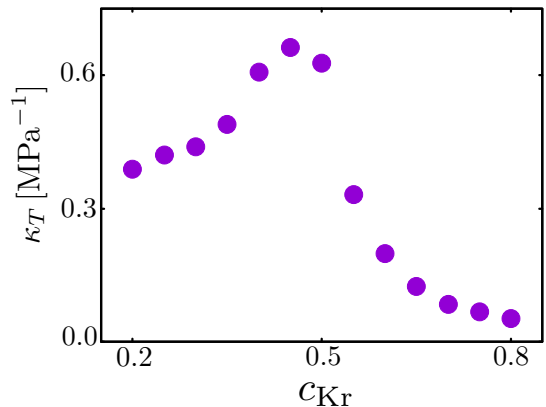}
\caption{Compressibility $\kappa_T$ for the mixture comprising argon and krypton as a function of the molar fraction $c_{\ce{Kr}}$ of krypton.}
\label{fig:kappaT}
\end{center}
\end{figure}

We present an example of $\Delta_{\rm mix} G$ determined from the molecular dynamics simulation for a mixture of argon and krypton at constant temperature and constant pressure. The mixture is modeled as three-dimensional Lennard--Jones liquids. 
We use the LAMMPS package in this demonstration. See Appendix \ref{ap:Ar-Kr} for details.
The molecules are packed in a rectangle container whose volume can fluctuate while keeping an aspect ratio of $21:5:5$ to fix the value of pressure.
The container is periodic in $y$ and $z$ directions whereas two boundary walls are set perpendicularly to the $x$ axis.  

We choose the values of temperature and pressure as $T=163.15\,{\rm K}$ and $p=4\,{\rm MPa}$,
at which liquid--vapor transition is observed with an increasing molar fraction  $c_{\ce{Kr}}=N_{\ce{Kr}}/N$ of the krypton \cite{nasrabad2004prediction}.
The total number of molecules is $N=500$, and 
the characteristics of the liquid--vapor transition are observed in numerical experiments.
Figures \ref{fig:Snap-RareGas} shows snapshots of the system's configuration after sufficient relaxation  for $c_{\ce{Kr}}=0.3$, $0.5$, and $0.65$. The volume differs greatly among the three values of $c_{\ce{Kr}}$. The number density at $c_{\ce{Kr}}=0.65$ is approximately $5$ times that at $c_{\ce{Kr}}=0.3$, and dense and dilute regions coexist at $c_{\ce{Kr}}=0.5$. Such behaviors clearly exhibit the characteristics of liquid--vapor transition. 
We also examine the compressibility $\kappa_{T} \equiv -\frac{1}{\bra V \ket}\pderf{\bra V \ket}{p}{T}$, which can be written as
\begin{align}
\kappa_{T}= \frac{\bra V^{2} \ket - \bra V \ket^{2}}{\kB T \bra V \ket }.
\end{align}
As shown in Fig.~\ref{fig:kappaT}, the compressibility decreases and approaches zero when $c_{\ce{Kr}}$ is larger than $0.55$, which indicates the behavior of liquid.
For $c_{\ce{Kr}}$ smaller than almost $0.35$, the mixture behaves as a gas,
with the compressibility being larger than that of liquid. 
We see that the compressibility grows more around $0.35<c_{\ce{Kr}}<0.55$. This is due to the coexistence of liquid and gas, for which the volume fluctuates largely. 
These observations are generally consistent with the results of a previous study on argon--krypton mixtures \cite{nasrabad2004prediction}.

With the above observations, we proceed to the determination of $\Delta_\mix G$.
Because the LAMMPS package does not contain the Jarzynski work relation, we 
calculate $\Delta_{(\#)}G$ and $\Delta_{\rm (ii)} G$ by 
the free-energy perturbation method \cite{zwanzig1954high} and substitute them into \eqref{e:Gmix-qs}.
In the calculations, we take the initial pure substance as being argon.

\begin{figure}
\begin{center}
\includegraphics[scale=0.64]{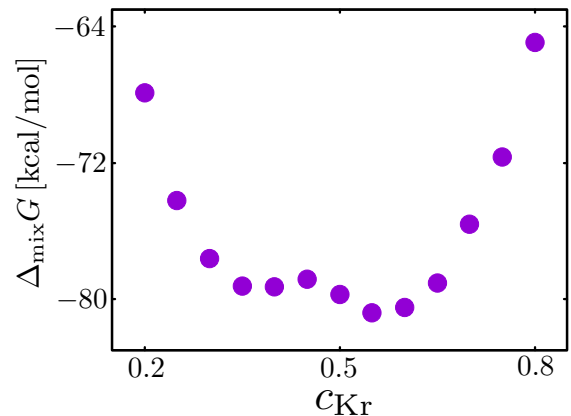}
\caption{$\Delta_\mix G$ for the binary mixture of argon and  krypton with $N=500$ determined using \eqref{e:Gmix-qs}.
Molecular dynamics simulations are performed with a unit time of $4.0 \, {\rm fs}$.  
The typical relaxation time is sufficiently shorter than $1.0 \, {\rm ns}$.
The alchemy operation producing the target mixture from the pure argon gas is divided into $20$ steps to apply the free-energy perturbation method for the calculations of $\Delta_{(\#)}G$ and $\Delta_{\rm (ii)}G$.
For each step, the system relaxes in $1.0 \, {\rm ns}$.  The number of samples is $25,000$. Here, 1 kcal/mol = 4.184 kJ/mol in SI unit.
}
\label{fig:Gmix-RareGas}
\end{center}
\end{figure}

The resulting mixing free energy $\Delta_\mix G$ is shown in Fig.~\ref{fig:Gmix-RareGas}.
The curve has a double-well shape and is convex upwards in the approximate range of $0.35<c_{\ce{Kr}}<0.55$,
which is consistent with the range in which the liquid--vapor coexistence is observed in the compressibility $\kappa_T$.
We thus conclude that the functional shape of $\Delta_\mix G$ well characterizes the liquid--vapor transition for the argon--krypton mixture.
Our formula, \eqref{e:Gmix}, \eqref{e:Gmix-N}, or \eqref{e:Gmix-qs}, actually works as a quantitative method for determining the mixing Gibbs free energy.

Note that $\ln N$ is  $1\%$ of $N$ at $N=500$ used in this demonstration, where
thermodynamic properties may deviate from those in the  thermodynamic limit. Indeed,  the upward convexity in $\Delta_\mix G$ is not expected in the thermodynamic limit from the second law of thermodynamics; i.e., the upward convex region should be flattened by increasing the system size $N$.
The coexistence states may become more unstable at small $N$ than in the thermodynamic limit owing to the enhanced fluctuations. Such finite size effects could be studied in terms of $\Delta_\mix G$ as an interesting future topic.

\section{Concluding remarks}\label{s:concluding}

We extended the scope of the work relation from macroscopic operations or single-molecule manipulations to microscopic operations producing a mixture from a pure substances.
To this end, we numerically derived the relation \eqref{e:distinguish}, in which the combinatorial factor $N!/n!(N-n)!$ was led from molecular dynamics simulations for classical molecule systems as shown in Figs. \ref{fig:CvsCstat}.
The free energy of the mixture determined  by the work relation \eqref{e:F_AB} or \eqref{e:G} is regarded as that measured in the standard reference by taking the free energy of the initial pure substance as the standard value in databases. The free energy in the standard reference makes it possible to compare thermodynamic properties among several mixtures.
We then proposed a variant of the work relation for determining the mixing Gibbs free energy characterizing  thermodynamic properties for the mixture. This variant is formulated as \eqref{e:Gmix}, \eqref{e:Gmix-N}, or \eqref{e:Gmix-qs} by combining two alchemical processes in Fig.~\ref{fig:G-op} and is connected to the excess chemical potential and activity coefficients for each substance in the mixture. 
We demonstrated the calculation of the mixing free energy for the mixture of argon and krypton, which clearly shows the characteristics of the liquid--vapor transition even in a small system of $N=500$.

We emphasize that formulas \eqref{e:Gmix}, \eqref{e:Gmix-N}, and \eqref{e:Gmix-qs} offer effective methods of numerically determining the mixing free energy. The advantages of the method are the generality of the subjected mixture, accessibility to the free energy, and low cost of the numerical computation.
Although we explored the method by limiting ourselves to a mixture of monoatomic molecules without electrical charges, the method would be applicable to various solutions with a general concentration, system size, species of molecules, and type of interaction.
For instance, the extension of the method to diatomic or polyatomic molecules is straightforward if the number of atoms for each molecule in the mixture is the same; e.g., a mixture of O$_2$ and N$_2$ or CO$_2$ and H$_2$O. 
To deal with a mixture comprising two species with the different numbers of atoms, 
we need to take care of the indistinguishability of atoms in the molecules.
We may avoid the difficulty by choosing the species of the initial pure substance as  the larger molecule. We
apply the alchemical process to cut the larger molecules into the same size as the smaller molecules and to change them into the target molecules. Respective alchemy tricks can be considered in accessing the mixing free energy for the respective target mixture.

Many solvation studies assume solutions to be dilute and apply the continuum limit approximation to the solvent \cite{skyner2015review}. 
Our method is free from such approximations and the reliability of the obtained mixing free energy depends on the reliability of the interaction potentials used, whose designs have been intensively studied for the development of molecular dynamics simulations \cite{sun1998compass}.
We here mention  methods of estimating the mixing free energy and activity coefficients. 
For ionic solutions of less ionic strength, the Debye-H\"{u}ckel theory and its extension are effective  \cite{Debye_1923, davies1938397}. 
For real solutions with general concentrations, heuristic approaches can be used to obtain an approximate perspective of the solutions. For instance, a method estimates activity coefficients using empirical models that require thermodynamic parameter inputs to be determined in other experiments.  Another method uses approximate partition functions by imposing simpler interaction potentials with which the partition function becomes accessible \cite{davies1938397, fredenslund1975group, klamt1995conductor}.
There, the reliability of the obtained values is rather obscure owing to the heuristic assumptions.

We next remark on a fundamental point raised by the numerical experiments of this paper. 
Our observations revealed the combinatorial factor as shown in \eqref{e:distinguish} and Figs.~\ref{fig:CvsCstat}.
This would be interpreted naturally as coming from the factorial $N!$ contained in the micro-canonical or canonical distribution, and may be universal over the choice of two substances. 
Because we adopted small system sizes, $N!$ was explicitly distinguished from another possible factor as $N^N$.
Let us recall that the factorial was initially introduced into classical statistical mechanics to satisfy the extensivity of free energy \cite{gibbs1875on, gibbs1902}. 
This was attributed to the indistinguishability of molecules, which was convincing owing to the consistency with quantum mechanics, although it led to the Gibbs paradox from a classical point of view.
The Gibbs paradox has been argued until now as a fundamental problem; e.g., the interpretation of the distinguishability from the fluctuation theorem \cite{murashita2017gibbs} and the ability to distinguish quantum systems \cite{holmes2020gibbs,yadin2021mixing}. 
We emphasize that our numerical experiments reveal the combinatorial factor only from the thermodynamic measurements of classical systems without making any assumption connected to quantum mechanics. 
This may be related to the fact that colloidal particle systems are accessible using a framework of statistical mechanics \cite{warren1998combinatorial}. 
The indistinguishability of molecules  may be derived by dealing with the right-hand side of \eqref{e:distinguish} theoretically for classical systems, which may shed new light on the Gibbs paradox.

Our numerical observations for isotopes with small $N$ reveal that the functional form of the mixing entropy is \eqref{e:Smix-iso} instead of \eqref{e:Smix-iso-notcombi}, which is consistent with the mixing entropy for ideal mixtures derived from statistical mechanics.
We here note that \eqref{e:S^id-apprx} does not only result from the combinatorial factor.
It is not $\ln \left[N!/n!(N-n)!\right]$ but requires the contribution of $\Delta_{\rm (i)}F$ from information thermodynamics. 
Combinatorial entropic effects due to the combination may be related to the stability of binding states of two biomolecules or absorption  states of small objects  \cite{liu2020combinatorial}, for which the informative contribution may play a role.

{\em Acknowledgment.---}  
The authors thank Takenobu Nakamura for valuable comments and technical information for numerical simulation, Shin-ichi Sasa, Kyousuke Tachi, Yuya Kai, Yohei Nakayama, and Minoru Kanega for fruitful discussions.
The computation in this work was done using the facilities of the Supercomputer Center, the Institute for Solid State Physics, The University of Tokyo. A. Y. is supported by Ibaraki University Fellowship Scholarship.
The present study was supported by KAKENHI (Nos. 17H01148, 19K03647, 20K20425).

\appendix

\section{Model}\label{ap:model}

We consider two-dimensional systems, where $N$ molecules are in a container of a rectangle box with dimensions of $L_x\times L_y$ and a periodic boundary condition in the $y$ direction.
The position of the $i$th molecule, $1\le i\le N$, is ${\bm r}_i=(x_i,y_i)$ with
$0\le x_i\le L_x$ and $0\le y_i<L_y$.
Two fixed walls are placed at $(0,y)$ and  $(L_x,y)$,
and  two movable membranes are at $(X_\subL(t),y)$ and $(X_\subR(t),y)$.
The mass and radius of the $i$th molecule are $m_i$ and  $r_0^i$,
 and the thickness of each wall and membrane is $\sigma_{\rm w}$. 
We set $\sigma_{\rm w}$ as almost vanishing compared with the radius  of the molecules.
We then define the system's Hamiltonian as
\begin{align}
&H(\Gamma;\bm{\alpha})=
\sum_{i=1}^N\frac{|{\bm p}_i|^2}{2m_i}+\Phi(\left\{\bm{r}_i\right\};\left\{r_0^i\right\},X_\subL,X_\subR,\lambda),
\label{e:H-v}
\end{align}
where $\lambda$ is a parameter that denotes the existence of the inserted membranes.
$\bm{\alpha}$ is the set of parameters in the Hamiltonian; i.e.,
\begin{align}
{\bm \alpha}=(\left\{m^i\right\},\left\{r_0^i\right\},X_\subL,X_\subR,\lambda).
\end{align}

We write the total potential of the system as
\begin{align}
&\Phi(\left\{\bm{r}_i\right\};\left\{r_0^i\right\},X_\subL,X_\subR,\lambda)=
\sum_{i=1}^N\sum_{j<i}\phi(|{\bm{r}_i}-{\bm{r}_j}|;r_0^i+r_0^j)\nonumber\\
&+
\sum_{i=1}^N\phi(x_i;r_0^i+\sigma_{\rm w})
+
\sum_{i=1}^N\phi(L_x-x_i;r_0^i+\sigma_{\rm w})\nonumber\\
&+\lambda\left[\sum_{i_\subL=1}^n\phi(X_\subR - x_{i_\subL};\sigma_{\rm w})
+\sum_{i_\subR=n+1}^{N}\phi(x_{i_\subR}-X_\subL ;\sigma_{\rm w})\right].
\label{e:Phi}
\end{align}
All the pair interactions between two objects are given by the Weeks--Chandler--Andersen (WCA) potential,
\begin{align}
\phi(r;\sigma)=
\begin{cases}
{\displaystyle 4\ep\left[\left(\frac{\sigma}{r}\right)^{12}-\left(\frac{\sigma}{r}\right)^{6}\right]+\ep,} & (r<2^\frac{1}{6}\sigma)\\
 0, &(r\ge 2^\frac{1}{6}\sigma)
\end{cases}
\end{align}
where $r$ is the distance for the interacting pair and $\sigma$ is the parameter given for respective pairs.

The first term of \eqref{e:Phi} is the interaction between two molecules, and the second and third terms are the interactions between the molecules and walls of the container.
The fourth and fifth terms of \eqref{e:Phi} correspond to the interaction between the semipermeable membranes and molecules.
$i_\subL$ and $i_\subR$ are the indices for labeling molecules observed in the left and right chambers, respectively.
Note that the membranes detect the molecules as points with mass. This was designed to reduce the excluded volume effect due to the insertion of the semipermeable membrane to zero.

We perform a molecular dynamics simulation with a Langevin thermostat having temperature $T$.
Each molecule evolves according to
\begin{align}
&\dot {\bm p}_i=-\pder{H}{{\bm r}_i}-\frac{\gamma(x_i)}{m_i} {\bm p}_i+\sqrt{2\gamma(x_i) \kB T}{{\bm \xi}_i}(t),\label{e:Langevin}
\end{align}
with $\dot {\bm r}_i={{\bm p}_i}/{m_i}$, where $\gamma(x_i)=1$ in the region $0<x_i<0.1 L_x$ and $0.9 L_x<x_i<L_x$ 
while $\gamma(x_i)=0$ in $0.1 L_x\le x_i\le 0.9 L_x$.
${\bm \xi}_i(t)=(\xi_i^x(t),\xi_i^y(t))$ is Gaussian white noise that satisfies $\sbkt{\xi_i^a(t)}=0$ and $\sbkt{\xi_i^a(t)\xi_j^b(t')}=\delta_{i,j}\delta_{a,b}\delta(t-t')$, where $a$ and $b$ are $x$ or $y$.
We take $\kB T=2.0\ep$, which is far above the Alder transition temperature.

\section{Protocols and works} \label{ap:protocol}

We take two examples of a mixture of two components.
The first example is a mixture of isotopes, where the two components have distinct mass.
One component is of $m$ while another is of $m+\Delta m$.
The second example is a mixture with molecules of different size, where the diameter of each component is parameterized by $r_0$ or $r_0+\Delta r_0$.
We take  $m=1$ and $r_0=2^{-1/6}$ in numerical calculations demonstrated in Appendices \ref{ap:rho_v} and \ref{ap:Fii}.

We first consider the protocol to produce the mixture of distinguishable molecules.
The $N$ molecules 
are totally indexed from $i=1$ to $N$ and relaxed to equilibrium with $\lambda=0$.
In the first example, we change $m$  with $r_0$ fixed as
\begin{align}
m_{i}(t)=m+\Delta m \frac{t}{\tau_{(\#)}},
\end{align}
for $1\le i\le n$. $\tau_{(\#)}$ is the operation time of the protocol.
According to formula \eqref{e:work}, 
the work required in the change of mass $m$ is  written as
\begin{align}
\hat W_{(\#)}^m=-\frac{\Delta m}{2\tau_{(\#)}}
\sum_{i=1}^n\int_0^{\tau_{(\#)}}dt~\left|\frac{{\bm p}_{i}(t)}{m(t)}\right|^2.
\end{align}
We can take $\tau_{(\#)}$ as being very short and even $\tau_{(\#)}\rightarrow 0$ when we use the Jarzynski relation \eqref{e:Jarzynski}.
In the second example, we change $r_0$ for the molecules with $m$ fixed as
\begin{align}
r_0^{i}(t)=r_0^i+\Delta r_0 \frac{t}{\tau_{(\#)}},
\end{align}
where  $1\le i\le n$. 
The required work is 
\begin{align}
\hat W_{(\#)}^{r_0}=&\frac{\Delta r_0}{\tau_{(\#)}}\sum_{i=1}^n \nonumber\\
&\int_0^{\tau_{(\#)}}\!\!\! dt~
\left.\pder{\Phi(\left\{{\bm r}_{i}(t)\right\};\left\{r_0^{i}\right\},0,1,0)}{r_0^{i}}\right|_{\{r_0^{i}\}=\{r_0^{i}(t)\}}.
\end{align}
When we consider a dilute fluid, we can take the period of operation as $\tau_{(\#)}\rightarrow 0$.

We next describe the protocol used to determine $\Delta_{\rm (i)} F$, $\Delta_{\rm (ii)} F$ and $\Delta_{\rm (iii)} F$.
For process (i), we put $N$ molecules in the container with $\lambda=0$ and relax the system in equilibrium.
At each moment $t$ after the relaxation,
we observe   the number of the molecules $n_v(\Gamma(t))$ 
in the region corresponding to the left chamber;  i.e., $x<v/L_y$.
From this observation, we construct the probability density $\rho_{v}(n)$.
We then obtain  $\Delta_{\rm (i)} F$ according to \eqref{e:F_i}.

In process (ii), we deal with  the system with $\lambda=1$ as it is separated by the wall. 
We choose an arbitrary value of $n$ and put $n$ molecules in the left chamber having volume $v$,
and $N-n$ molecules in the right chamber having volume $V-v$.
In this paper, we choose $n=N v/V$ 
because the left and right systems become almost equivalent.
We label molecules in the left and right chambers as $i_\subL$ and $i_\subR$, respectively,
where $1\le i_\subL\le n$ and $n+1\le i_\subR\le N$.
We relax this combined system to equilibrium.
Note that the procedures up to here are common for all mixtures that we want to examine.
We then start alchemical process (ii).
In the first example, we set the initial mass of all molecules as $m_i=m$.
We change the mass of the left $n$ molecules while fixing that of the right molecules as
\begin{align}
m_{i_\subL}(t)=m+\Delta m \frac{t}{\tau_{\rm (ii)}}, \qquad  m_{i_\subR}(t)=m,
\label{e:m-protocol}
\end{align}
for $0\le t\le \tau_{\rm (ii)}$. 
In $t\ge \tau_{\rm (ii)}$, the masses are fixed as $m+\Delta m$ and $m$ in left and right chambers, respectively.
The  work required for this change is 
\begin{align}
\hat W_{\rm (ii)}^m=-\frac{\Delta m}{2\tau_{\rm (ii)}}
\sum_{i_\subL=1}^n\int_0^{\tau_{\rm (ii)}}dt~\left|\frac{{\bm p}_{i_\subL}(t)}{m_{i_\subL}(t)}\right|^2.
\end{align}

In the second example, we set the initial  radius of all molecules as $r_0$ and then change the radius in the left chamber
as
\begin{align}
r_0^{i_\subL}(t)=r_0+\Delta r_0 \frac{t}{\tau_{\rm (ii)}}, \qquad  r_0^{i_\subR}(t)=r_0,
\label{e:r0-protocol}
\end{align}
for $0\le t\le \tau_{\rm (ii)}$, and then fix the radiuses.
The required work is
\begin{align}
\hat W_{\rm (ii)}^{r_0}=
&\frac{\Delta r_0}{\tau_{\rm (ii)}}\sum_{i_\subL=1}^n
\nonumber\\
&\int_0^{\tau_{\rm (ii)}}\!\!\! dt\left.\pder{\Phi(\left\{{\bm r}_{i}(t)\right\};\left\{r_0^{i}\right\},\frac{v}{L_y},\frac{v}{L_y},1)}{r_0^{i_\subL}}\right|_{\{r_0^{i_\subL}\}=\{r_0^{i_\subL}(t)\}}.
\end{align}

We equilibrate the system for an interval $t_{\rm r}$ sufficiently longer than the system's relaxation time.
We then proceed to protocol (iii), which is the hardest process in the computation.
Note that the pressure will be different between the left and  the right chambers, especially in the second example.
Such a difference may be a cause of irreversibility; however, 
this does not matter in principle for the use of the Jarzynski work relation.

We start to move the two membranes at time $t_1=\tau_{\rm (ii)}+t_{\rm r}$, which is expressed as
\begin{align}
&X_\subL(t)=\frac{v}{L_y}\left(1-\frac{t-t_1}{\tau_{\rm (iii)}}\right),\\
&X_\subR(t)=\frac{V}{L_y}-\frac{V-v}{L_y}\left(1-\frac{t-t_1}{\tau_{\rm (iii)}}\right)
\end{align}
for $t_1<t<t_1+\tau_{\rm (iii)}$,
in which the operation time $\tau_{\rm (iii)}$ is as long as
\begin{align}
\tau_{\rm (iii)}\sim \max\left(\frac{v}{r_0^{i_\subL}(t_1)}, \frac{V-v}{r_0^{i_\subR}}\right)
\end{align}
to avoid numerical errors and/or divergence.
The works along this protocol are
\begin{align}
&\hat W_{\rm (iii)}^\subL=
-\frac{v}{L_y}\frac{1}{\tau_{\rm (iii)}}\nonumber\\
&\int_{t_1}^{t_1+\tau_{\rm (iii)}}\!\!\! dt~
\left.\pder{\Phi(\left\{{\bm r}_{i}(t)\right\};\left\{r_0^{i}(t_1)\right\},X_\subL,X_\subR(t),1)}{X_\subL}\right|_{X_\subL=X_\subL(t)}
\end{align}
and
\begin{align}
&\hat W_{\rm (iii)}^\subR=
\frac{V-v}{L_y}\frac{1}{\tau_{\rm (iii)}}\nonumber\\
&\int_{t_1}^{t_1+\tau_{\rm (ii)}}\!\!\! dt~
\left.\pder{\Phi(\left\{{\bm r}_{i}(t)\right\};\left\{r_0^{i}(t_1)\right\},X_\subL(t),X_\subR,1)}{X_\subR}\right|_{X_\subR=X_\subR(t)}.
\end{align}

\section{General estimate of $\Delta_{\rm (i)} F$} \label{ap:rho_v}

\begin{figure}
\begin{center}
\includegraphics[scale=0.55]{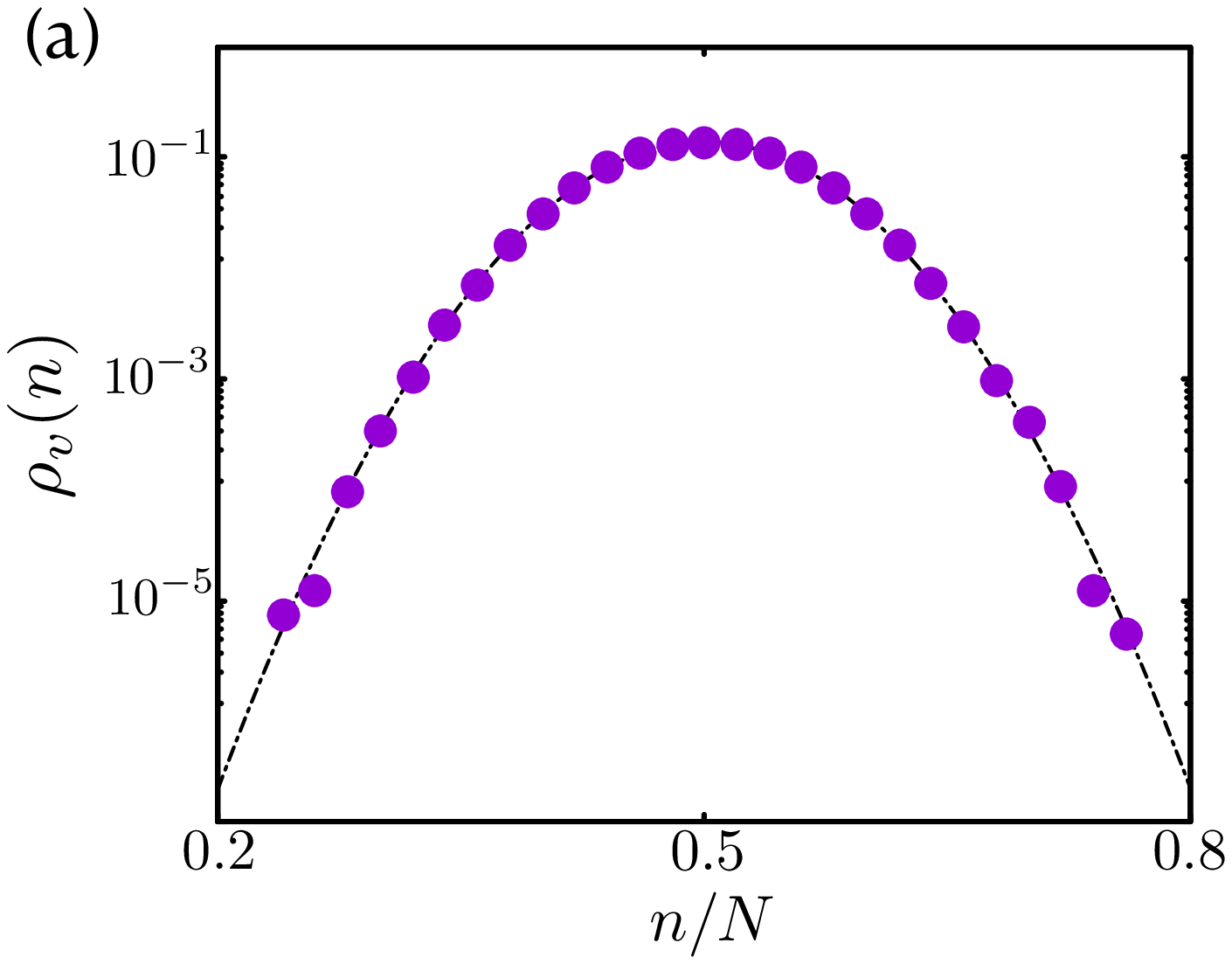}
\includegraphics[scale=0.55]{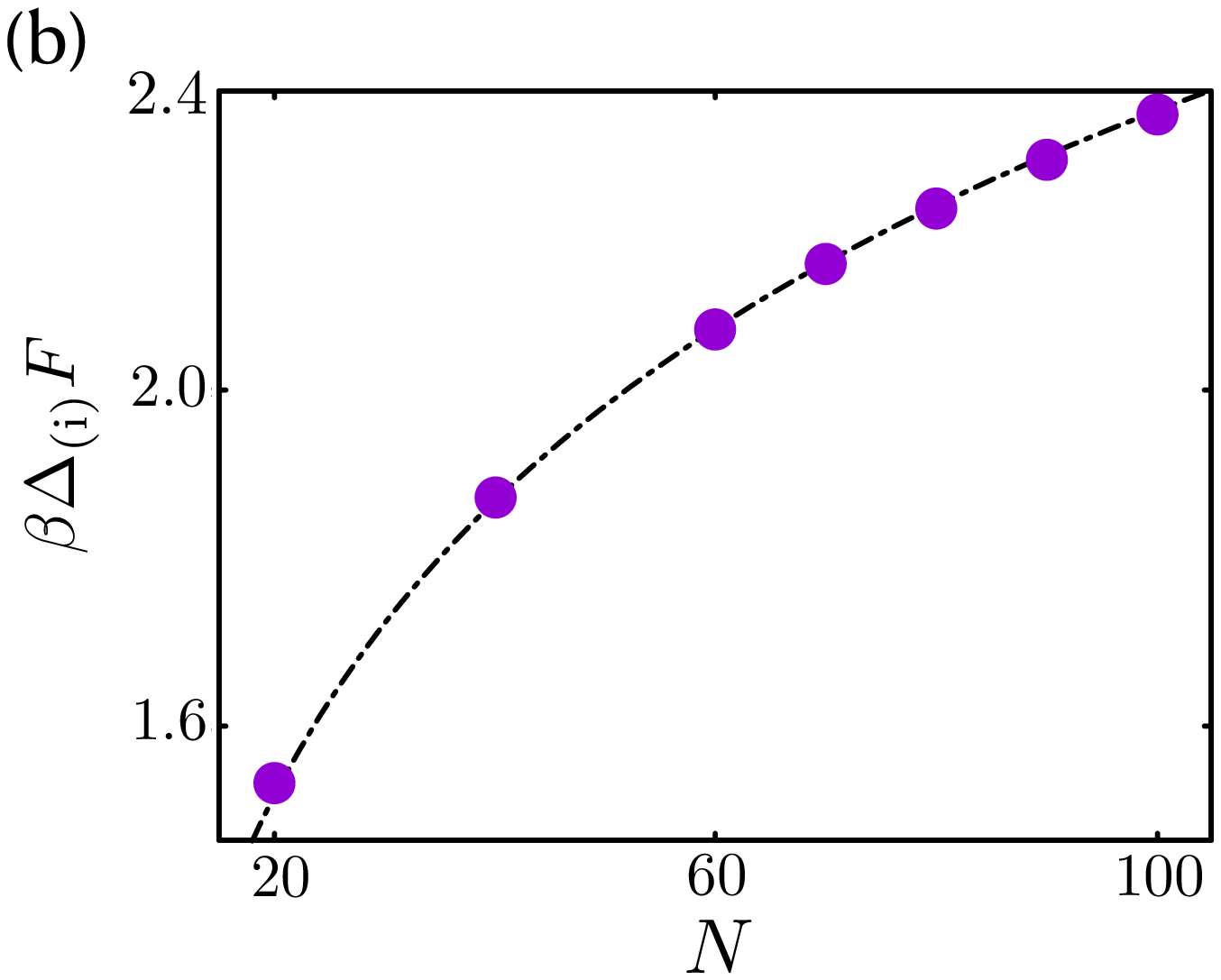}
\caption{(a) Distribution of the number of molecules $\rho_{v}(n)$ for $N=50$.
  The line indicates the Gaussian distribution
  with $\sbkt{n}=N/2$ and $\sigma(n)=0.42\sqrt{N}$.
(b) $\Delta_{\rm (i)} F$ resulting from the numerically determined $\rho_v(n)$ at $n=Nv/V$ for $20\le N\le 100$. The line indicates $\beta\Delta_{\rm (i)} F=\frac{1}{2}\ln N$ as \eqref{e:F_i-gauss}.
To obtain each point, $\rho_v(n=Nv/V)$ is determined from $50,000$ samples for $V=30N r_0^2$ and $v=V/2$
in both (a) and (b). 
}
\label{fig:Pn}
\end{center}
\end{figure}

Figure ~\ref{fig:Pn}(a) shows an example of the distribution $\rho_v(n)$ for $v=V/2$ 
 when $N=50$.
It clearly shows that $\rho_v(n)$ is approximated well by a Gaussian distribution exhibited by a line.
We calculate $\Delta_{\rm (i)} F$ by \eqref{e:F_i} from numerically determined $\rho_v(\sbkt{n})$.
As shown in Fig.~\ref{fig:Pn}(b) for $20\le N\le 100$, $\Delta_{\rm (i)} F$ exhibits a logarithm of $N$, 
which will be common over species of the initial substance, as explained below. 

Once we choose the values of $v$, $V$, and $N$, we naturally expect the mean number of the molecules in the left chamber to be $\sbkt{n}= N v/V$. We here assume $v=O(V)$ and $V-v=O(V)$.
The probability distribution $\rho_v(n)$ is generally written as
\begin{align}
&\rho_{v}(n)=\frac{1}{C_v(N)}\exp\left[-N\sum_{k=2}^\infty \frac{a_k}{k!}\left(\frac{n}{N}-\frac{\sbkt{n}}{N}\right)^k\right],\\
&C_v(N)=N \int_0^1 dc~ \exp\left[-N\sum_{k=2}^\infty \frac{a_k}{k!}\left(c-\sbkt{c}\right)^k\right],
\end{align}
where $a_k$ is a constant of $O(N^0)$ and $c=n/N$.
Substituting  $n=\sbkt{n}$ into the above general form, we have
\begin{align}
\ln\rho_v(\sbkt{n})=-\ln C_v(N).
\end{align}
A standard procedure for large $N$ leads to an estimate as
\begin{align}
C_v(N)=\sqrt{N}\left(\sqrt{\frac{2\pi}{a_2}}+o(N^{-\frac{1}{2}})\right),
\end{align}
which yields
\begin{align}
\ln\rho_v(\sbkt{n})=-\frac{1}{2}\ln N+o(\ln N).
\label{e:Gaussian}
\end{align}
Therefore, especially for $n=\bra n \ket = Nv/V$, (15) is rewritten as
\begin{align} 
\beta{\Delta_{\rm (i)}F}=\frac{1}{2}\ln N+o(\ln N)
\label{e:F_i-gauss}
\end{align}
from (C5). Note that this formula holds universally for any mixture.
The numerical results are presented in Fig.~\ref{fig:Pn}(b), which shows a good agreement with \eqref{e:F_i-gauss} depicted as a line.

\section{Numerical results on respective free-energy changes}\label{ap:Fii}

\begin{figure}[tb]
\begin{center}
\includegraphics[scale=0.6]{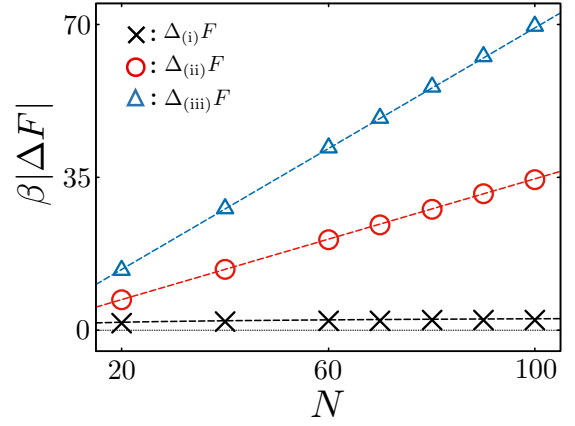}
\caption{Free-energy changes $\Delta_{\rm (i)} F$, $\Delta_{\rm (ii)} F$, and $\Delta_{\rm (iii)} F$ for the change in mass $m\rightarrow 2m$ as a function of $N$.
The volume $v$ for the left chamber is chosen as $v = Vn/N$, where $V=6N r_0^2$ and $n/N=0.5$. 
Lines are estimates of \eqref{e:stat1}, \eqref{e:stat2}, and \eqref{e:stat3} from statistical mechanics.
}
\label{fig:deltaF_m}
\end{center}
\end{figure}

\begin{table}[tb]
\begin{center}
\begin{tabular}{|c|c|c|c|} \hline
  process & $a$  & $b$ & operation time \\ \hline \hline
    (ii) & $-0.346 \pm 0.001$  & $0.003 \pm 0.018$ & 400$\tau_{\rm MD}$\\ \hline
    (iii) & $-0.700 \pm 0.001$  & $0.069 \pm 0.019$ & 30000$\tau_{\rm MD}$\\ \hline
    $(\#)$ & $-0.349 \pm 0.005$  & $0.033 \pm 0.010$ & 400$\tau_{\rm MD}$\\ \hline
  \end{tabular}
  \end{center}
  \caption{Values of fitting parameters $a$ and $b$ in \eqref{e:fitting} 
   when changing the mass as $m\rightarrow 2m$ for $n$ molecules. $n=0.5 N$,  $V=6N r_0^2$ and $v=0.5V$.
   Operation times $\tau_{\rm (ii)}$ and $\tau_{\rm (iii)}$ are depicted in the last columns.}
   \label{tab:deltaF_m}
\end{table}

We here demonstrate numerical estimates of respective free-energy changes to clarify their $N$ dependence.

We focus on small values of $N$ as $N\le 100$,
where $\ln N!$ largely deviates from $N\ln N$. The difference is estimated using Stirling's formula as
\begin{align}
\ln N!-N\ln N=-N+\frac{1}{2}\ln N+o(\ln N).
\end{align}
From this formula with $n=O(N)$, we have 
\begin{align}
&\ln \frac{N!}{n!(N-n)!}-\left[n\ln \frac{n}{N}-(N-n)\ln\frac{N-n}{N}\right]\nonumber\\
&=\frac{1}{2}\ln\frac{N}{n(N-n)}+o(\ln N),
\end{align}
whose right-hand side is $O(\ln N)$ and ignored at sufficiently large $N$.
This indicates that identifying the contribution of $O(\ln N)$ for each free-energy difference makes the finite size effect on $N$ clear.
We therefore make numerical estimates up to $O(\ln N)$ for each free-energy difference.
We fit the numerical results in the functional form as
\begin{align}
\beta{\Delta_{\rm (ii, iii, \#)}F}=a N+b\ln N,
\label{e:fitting}
\end{align} 
where the first term on the right-hand side corresponds to the extensive contribution remaining in the thermodynamic limit.
The second term is important to the purpose of this paper.

For numerical estimates of $\Delta_{\rm (ii)} F$ and $\Delta_{\rm (iii)} F$, 
we choose $n/N=v/V=0.5$ and
set $V=6Nr_0^2$ for the mixture of isotopes, whereas $V=30 N r_0^2$ for the mixture of the different size molecules.
We calculate $5000$ samples for each protocol.

Figure \ref{fig:deltaF_m} shows the respective free-energy changes 
in the protocol $m\rightarrow 2m$, where the mixture comprises isotopes.
The operation times are $\tau_{\rm (ii)}=400 \tau_{\rm MD}$ and $\tau_{\rm (iii)}=30000\tau_{\rm MD}$, where $\tau_{\rm MD}\equiv 2r_0 \sqrt{m/\ep}$.
As seen, both $\Delta_{\rm (ii)} F$ and $\Delta_{\rm (iii)}F$ increase linearly with $N$,
which become far superior to $\Delta_{\rm (i)} F$ at $N=100$. 
The fitting parameters $a$ and $b$ are summarized in Table \ref{tab:deltaF_m}.
We find that the coefficient $b$ is sufficiently small to ignore the contribution of $O(\ln N)$.
Thus, the contribution of $O(\ln N)$ in $\Delta F$ comes only from $\Delta_{\rm (i)}F$,
which indicates the importance of $\Delta_{\rm (i)} F$ to estimate free energy in the finite-size systems.

When the isotopes are an ideal gas, we can directly calculate the respective free-energy change using statistical mechanics. 
We derive in Appendix \ref{ap:statmech}
\begin{align}
&\beta\Delta_{\rm (i)}F=\frac{1}{2}\ln N+o(\ln N),\label{e:stat1}\\
&\beta\Delta_{\rm (ii)}F=-\frac{N}{2}\ln 2+o(\ln N),\label{e:stat2}\\
&\beta\Delta_{\rm (iii)}F=-N\ln 2+o(\ln N).\label{e:stat3}
\end{align}
We show these estimates as the lines in Fig.~\ref{fig:deltaF_m}.
Even though we adopt a finite radius with  $r_0\neq 0$,
numerical results fit well to these theoretical results for the ideal isotopes.

\begin{figure}[tb]
\begin{center}
\includegraphics[scale=0.6]{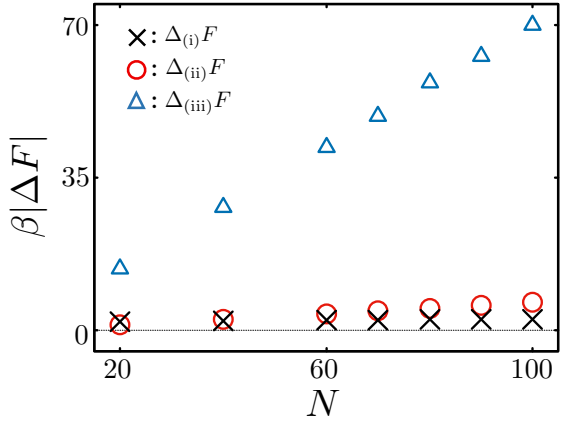}
\caption{Free-energy changes $\Delta_{\rm (i)} F$, $\Delta_{\rm (ii)} F$, and $\Delta_{\rm (iii)} F$ for the change in radius $r_0\rightarrow 2r_0$ as a function of $N$.
The volume $v$ for the left chamber is chosen as $v = Vn/N$, where $V=30N r_0^2$ and $n/N=0.5$.  
}
\label{fig:deltaF_r}
\end{center}
\end{figure}

Figure \ref{fig:deltaF_r} displays numerical results  for the protocol $r_0\rightarrow 2r_0$.
We remark $\Delta_{\rm (ii)}F \neq \Delta_{(\#)}F$ as demonstrated in Table \ref{tab:deltaF_r}. This is an important difference from the isotopes with $\Delta_{\rm (ii)}F=\Delta_{(\#)}F$.
The difference between $\Delta_{\rm (ii)}F$ and $\Delta_{(\#)}F$ may indicate that the deviation of the mixture from the ideal one and characterize the nontrivial thermodynamic properties of the mixture.

\begin{table}[tb]
\begin{center}
\begin{tabular}{|c|c|c|c|c|} \hline
  process & $a$  & $b$ & operation time \\ \hline \hline
    (ii) & $0.066  \pm 0.001$  & $-0.060 \pm 0.004$ & $2050 \tau_{\rm MD}$ \\ \hline
    (iii) & $-0.700 \pm 0.007$  & $-0.046 \pm 0.120$ & $15000 \tau_{\rm MD}$\\ \hline
   $(\#)$ & $0.058(6) \pm 0.000(1)$ & $-0.040\pm 0.002$ & $2050 \tau_{\rm MD}$\\ \hline
  \end{tabular}
  \end{center}
  \caption{
  Values of fitting parameters $a$ and $b$ in \eqref{e:fitting} 
   when changing the radius as $r_0\rightarrow 2r_0$ for $n$ molecules. $n=0.5 N$,  $V=6N r_0^2$ and $v=0.5V$.
     Operation times $\tau_{\rm (ii)}$ and $\tau_{\rm (iii)}$ are depicted in the last columns.}
     \label{tab:deltaF_r}
\end{table}

\section{Free-energy changes for ideal solutions of isotopes derived from statistical mechanics}\label{ap:statmech}

The free energy for the solution of two ideal isotopes can be calculated theoretically according to statistical mechanics.
The partition function for pure substance $\subA$ or $\subB$ is calculated as
\begin{align}
&Z_\subA(V,N) 
= \frac{( 2\pi \kB T m V )^{N}}{N!} ,\\
&Z_\subB(V,N) 
= \frac{( 2\pi \kB T (m+\Delta m) V )^{N}}{N!} ,
\end{align}
whereas that of the mixture of $\subA$ and $\subB$ is
\begin{align}
Z_\subAB(V,n,N-n)
= \frac{( 2\pi \kB T  V )^{N}m^{n}(m+\Delta m )^{N-n}}{n!(N-n)!}.
\end{align}
Since $\beta F=-\ln Z$,  we have the respective differences of the free energy as
\begin{align}
&\beta\Delta F
= - n \ln \frac{m+\Delta m}{m} - \ln \frac{N!}{n!(N-n)!},
\label{e:deltaF-stat}
\\
&\beta\Delta_{\rm (ii)} F = -n\ln \frac{m+\Delta m}{m},
\label{e:F_ii-stat}
\\
&\beta{\Delta_{\rm (iii)}F}=n\ln\frac{n}{N}+(N-n)\ln\frac{N-n}{N}.
\label{e:F_iii-stat}
\end{align}
Note that the right-hand side of \eqref{e:F_iii-stat} is nothing but the mixing entropy for the ideal solution $-\kB\Delta_\mix S^{\id}$.
This is because process (iii) for the  isotope mixture does not change the internal energy and pressure of the system, which leads to $\Delta_{\rm (iii)}F=-T\Delta_\mix S^{\id}$.

Recalling that $\Delta_{\rm (i)}F=\Delta F-\Delta_{\rm (ii)}F-\Delta_{\rm (iii)}F$, \eqref{e:deltaF-stat}, \eqref{e:F_ii-stat}, and \eqref{e:F_iii-stat} yield
\begin{align}
\beta{\Delta_{\rm (i)}F}=- \ln \frac{N!}{n!(N-n)!}-n\ln\frac{n}{N}-(N-n)\ln\frac{N-n}{N}.
\label{e:F_i-stat}
\end{align}
\eqref{e:F_i-stat} agrees with the estimate \eqref{e:rho_v-apprx} or \eqref{e:Gaussian}.
Applying Stirling's formula to \eqref{e:F_i-stat}, we obtain 
\begin{align}
&\beta\Delta_{\rm (i)}F=\frac{1}{2} \ln N + \frac{1}{2} \ln \frac{2 \pi n(N-n)}{N^{2}}+ o(N^{0}).
\end{align}
This is consistent with \eqref{e:rho_v-apprx} because the second term of the right-hand side is $o(\ln N)$
when $n/N$ and $(N-n)/N$ are $O(N^0)$.

Let us calculate $\Delta_{\rm (i)} F$ directly. 
The probability that one molecule of ideal gas exists in a region of volume $v$ is $v/V$,
and the probability of finding $n$ molecules in the region of volume $v$ is thus given by a binomial distribution,
\begin{align}
\rho_{v}(n) = \left( \frac{v}{V} \right)^{n} \left( \frac{V-v}{V} \right)^{N-n} \frac{N!}{n!(N-n)!}.
\end{align}
Substituting this form of $\rho_v(n)$ with $n=Nv/V$ into \eqref{e:F_i}, we obtain \eqref{e:F_i-stat}. 
This agreement  convinces us of the validity of \eqref{e:F_i} as the formula of $\Delta_{\rm (i)}F$.

\section{Model for the mixture of argon and krypton}\label{ap:Ar-Kr}

The interaction of any two molecules, argon or krypton, is given by the Lennard--Jones potential, 
\begin{align}
\phi(r;\ep,\sigma)=
\begin{cases}
{\displaystyle 4\ep\left[\left(\frac{\sigma}{r}\right)^{12}-\left(\frac{\sigma}{r}\right)^{6}\right],} & (r<r_{\rm c})\\
 0,  &(r\ge r_{\rm c})
\end{cases}
\end{align}
where $r_{\rm c}$ is the cutoff length.
The parameters of $\phi$ are set as reported in \cite{oh2013modified};
for the argon pair, $\sigma_{\ce{Ar}}=3.401\,{\rm \AA}$ and $\ep_{\ce{Ar}}=0.2321\,{\rm kcal/mol}$,
whereas $\sigma_{\ce{Kr}}=3.601\,{\rm \AA}$, $\ep_{\ce{Kr}}=0.3270\,{\rm kcal/mol}$ for the krypton pair. Here, kcal is defined by the thermochemical calorie as 1 kcal/mol = 4.184 kJ/mol.
For  the pair of argon and krypton, the Lorentz--Berthelot law is assumed as is usual for the Lennard--Jones binary mixture \cite{lorentz1881, berthelot1889}, $\sigma_{\ce{Ar}\ce{Kr}}=(\sigma_{\ce{Ar}} + \sigma_{\ce{Kr}})/2= 3.501\,{\rm \AA}$ and $\ep_{\ce{Ar}\ce{Kr}} = \sqrt{\ep_{\ce{Ar}}  \ep_{\ce{Kr}} } = 0.2755\,{\rm kcal/mol}$.
We set the cutoff length as $r_{\rm c}=3\sigma_{\ce{Kr}}$.
The masses of argon and krypton are $m_{\ce{Ar}}=39.95\, {\rm g/mol}$ and $m_{\ce{Kr}}=83.80 \, {\rm g/mol}$.

The molecules are packed in a rectangular container, which is periodic in $y$ and $z$ directions whereas two soft-core walls with $\sigma_{\rm w}=\sigma_{\ce{Ar}}/2$ are set as they are perpendicular to the $x$ axis.
The aspect ratio of the container is kept at $21:5:5$.

The numerical simulation is performed at constant temperature and constant pressure using the LAMMPS molecular dynamics package. The temperature and pressure are controlled by the Nose--Hoover chain and Martyna--Tobias--Klein barostat, respectively \cite{martyna1994constant}.

\bibliography{draft10}

\end{document}